\documentclass[reqno,twoside,a4paper,11pt]{article}
\usepackage[utf8]{inputenc}
\usepackage[english]{babel}
\usepackage{color}
\usepackage{tocloft}
\usepackage{tabularx,hhline}
\usepackage{enumitem}

\usepackage[
  hmarginratio={1:1},     
  vmarginratio={1:1},     
  textwidth=16cm,        
  textheight=24cm,
  heightrounded,          
]{geometry}

\usepackage{amsmath,amsthm,amssymb,accents}
\usepackage{mathrsfs, mathtools}
\usepackage{dsfont}

\usepackage[all]{xy} 
\input xy 
\xyoption{all}
\xyoption{2cell} 
\xyoption{v2}
\SelectTips{cm}{11} 

\usepackage{subdepth}

\usepackage[hidelinks]{hyperref}

\usepackage[numbers,sort&compress]{natbib}

\usepackage[mathscr]{euscript}

\newcommand{\red}{\mathrm{W}}
\newcommand{\rmpar}{\mathrm{par}}
\newcommand{\pt}{\mathrm{pt}}
\newcommand{\rmH}{\mathrm{H}}
\newcommand{\rmD}{\mathrm{D}}
\newcommand{\rmL}{\mathrm{L}}

\newcommand{\dR}{\mathrm{dR}}
\newcommand{\dd}{\mathrm{d}}
\newcommand{\DD}{\mathrm{DD}}
\newcommand{\rmR}{\mathrm{R}}

\newcommand{\Diff}{\mathrm{Diff}}

\newcommand{\pre}{\mathrm{pre}}
\newcommand{\rmt}{\mathrm{t}}
\newcommand{\rmh}{\mathrm{h}}

\newcommand{\scH}{\mathscr{H}}

\newcommand{\scC}{\mathscr{C}}

\newcommand{\scD}{\mathscr{D}}
\newcommand{\scB}{\mathscr{B}}

\newcommand{\CI}{\mathcal{I}}
\newcommand{\CH}{\mathcal{H}}

\newcommand{\CL}{\mathcal{L}}
\newcommand{\CP}{\mathcal{P}}
\newcommand{\CU}{\mathcal{U}}
\newcommand{\CD}{\mathcal{D}}

\newcommand{\CN}{\mathcal{N}}
\newcommand{\CS}{\mathcal{S}}
\newcommand{\CO}{\mathcal{O}}

\newcommand{\sfG}{\mathsf{G}}

\newcommand{\sfP}{\mathsf{P}}

\newcommand{\sfU}{\mathsf{U}}
\newcommand{\sfS}{\mathsf{S}}

\newcommand{\sfB}{\mathsf{B}}

\newcommand{\Mat}{\mathsf{Mat}}

\newcommand{\FC}{\mathbb{C}}
\newcommand{\FL}{\mathbb{L}}
\newcommand{\FR}{\mathbb{R}}
\newcommand{\RZ}{\mathbb{Z}}

\newcommand{\NN}{\mathbb{N}}

\newcommand{\frh}{\mathfrak{h}}
\newcommand{\frg}{\mathfrak{g}}
\newcommand{\fru}{\mathfrak{u}}

\newcommand{\Vect}{\mathscr{V}\mathrm{ect}}

\newcommand{\HVBdl}{\mathscr{HVB}\hspace{-0.01cm}\mathrm{dl}}
\newcommand{\HLBdl}{\mathscr{HL}\hspace{-0.02cm}\mathscr{B}\mathrm{dl}}
\newcommand{\Sh}{\mathscr{S}\hspace{-0.01cm}\mathrm{h}}
\newcommand{\BGrb}{\mathscr{B}\hspace{-.025cm}\mathscr{G}\mathrm{rb}}
\newcommand{\Hilb}{\mathscr{H}\mathrm{ilb}}

\newcommand{\disc}{\scD\mathrm{isc}}

\newcommand{\smint}{\textstyle\int}

\newcommand{\arisom}{\overset{\cong}{\longrightarrow}}
\newcommand{\Arisom}{\overset{\cong}{\Longrightarrow}}

\newcommand{\invs}[1]{{#1}\hspace{-0.05cm}^\times}

\newcommand{\iu}{\mathrm{i}}
\newcommand{\tr}{\mathrm{tr}}

\newcommand{\<}{\langle}
\renewcommand{\>}{\rangle}

\newenvironment{myitemize}{\begin{itemize}[itemsep=0.cm]}{\end{itemize}}

\newtheoremstyle{thm} 								
{0.15cm}   					
{0.cm}   	 				
{\itshape} 	
{}         				
{\bfseries}				
{}        				
{0.2cm} 					
{}         				

\newtheoremstyle{rmk} 								
{0.15cm}   					
{0.cm}   	 				
{} 	
{}         				
{\bfseries}				
{}        				
{0.2cm} 					
{}         				

\theoremstyle{thm}

\theoremstyle{rmk}

\numberwithin{equation}{section}

\makeatletter
\newlength{\@thlabel@width}%
\newcommand{\thmenumhspace}{\settowidth{\@thlabel@width}{(1)}\sbox{\@labels}{\unhbox\@labels\hspace{\dimexpr-\leftmargin+\labelsep+\@thlabel@width-\itemindent}}}
\makeatother

\begin{document}

\begin{flushright}
\small
EMPG--16--19
\end{flushright}

\vskip .5cm

%
%


\begin{center}
	\textbf{\LARGE{Fluxes, bundle gerbes and 2-Hilbert spaces}}\\
	\vspace{0.5cm}
	\Large Severin Bunk and Richard J. Szabo
\end{center}

\begin{abstract} 
\noindent
We elaborate on the construction of a prequantum 2-Hilbert space from a bundle gerbe over a 2-plectic manifold,
providing the first steps in a program of higher geometric quantisation
of closed strings in flux compactifications and of M5-branes in
$C$-fields. We review in detail the construction of the 2-category
of bundle gerbes, and introduce the higher geometrical structures necessary to
turn their categories of sections into 2-Hilbert spaces. We work out several
explicit examples of 2-Hilbert spaces in the context of closed strings
and M5-branes on flat space. We also work out the prequantum 2-Hilbert space associated to
an M-theory lift of closed strings described by an asymmetric cyclic orbifold of the
$\sfS\sfU(2)$ WZW model, providing an example of
sections of a torsion gerbe on a curved background. We describe the dimensional reduction of
M-theory to string theory in these settings as a map from 2-isomorphism
classes of sections of bundle gerbes to sections of corresponding line
bundles, which is compatible with the respective monoidal structures
and module actions.
\end{abstract}



\tableofcontents

\bigskip

\section{Introduction and summary}

The geometry of fluxes in string theory and M-theory has over the
years spawned many new directions in classical geometry and
quantisation, see e.g.~\cite{Aschieri-Szabo--NonAsso,BDL--NonAsso,Blumenhagen:2010hj,Chu:2009iv,Lust:2010iy,Mylonas:2012pg}
In many of these developments the need for higher
geometrical structures together with higher notions of quantisation
is becoming ever more prominent (see for instance~\cite{SS--Groupoids_loops_GeoQuan} for a survey of the literature and further references).

Physical theories are usually formulated at the classical level using geometry and variational principles. Quantisation can then be understood as a mathematical construction applied to the classical data that describes a quantum version of the physical theory in question. Seeking new formalisms for quantisation is crucial in understanding how to extract physical properties from a mathematical quantum theory, and to properly explore the passage from classical mechanics to quantum mechanics. In particular, beyond its intrinsic interest in string theory and M-theory, seeking the appropriate receptacles for higher versions of quantisation can offer interesting new perspectives on quantum theory itself and a move towards a more systematic theory of quantisation.

Evidence that such higher quantisations are possible has been provided in a number of works, including~\cite{Nuiten--Thesis,Bongers--Thesis,FSS--Higher_stacky_CS,FRS--Higher_U1_gerbe_conns,Brylinski}.
In this series of works, higher prequantisation has been investigated not just for the 2-plectic case which is central to this paper, but for any degree $n \geq 2$ of the higher symplectic form.
In the present paper, we make some of the abstract objects in these references very explicit, using the more differential geometric language of bundle gerbes rather than the powerful but abstract language of simplicial sheaves.
At the same time, using the 2-categorical theory developed in~\cite{Waldorf--More_morphisms}, we are able to use a more global framework as compared to~\cite{Rogers--Thesis}, and develop further the ideas about using twisted vector bundles as quantum states outlined there.
To our knowledge, the notion of \emph{sections of a bundle gerbe} $\CL$ in terms of morphisms from a trivial bundle gerbe to $\CL$ has not yet appeared in the literature, though it has been around as folklore, see~\cite{Waldorf--Sections_on_mathoverflow}.
It has also been pointed out there that the resulting categories are Kapranov-Voevodsky 2-vector spaces, and we extend this towards higher prequantisation by showing that these are actually 2-Hilbert spaces (cf.~Section~\ref{sect:2Hspace_of_BGrb}).

The basic problem of quantisation begins with a symplectic manifold $M$, with symplectic form $\omega$, which is the classical phase space of a given physical system. Quantisation can then be loosely regarded as a procedure for constructing a ``quantum
version'' of symplectic geometry. In most approaches to quantisation, the first step (either explicitly or implicitly) is to pick the additional structure of a prequantum line bundle $L\to M$, which is a hermitean line bundle with connection whose magnetic flux is proportional to the 2-form $\omega$. In this paper we deal with geometric quantisation~\cite{Kostant-1970aa,Kirillov2001}, which is one of the best established quantisation procedures. We review its basic features in Section~\ref{sect:GeoQuan_review}, and then confront it with string theory and M-theory in Section~\ref{sect:strings_and_tautological_gerbe}: We explain why geometric quantisation is not appropriate for the quantisation of closed strings in flux compactifications or of M5-branes in $C$-field backgrounds of M-theory, focusing on the particular cases that involve replacing the symplectic form $\omega$ with a suitable 3-form. As we discuss in Section~\ref{sect:Bundle_gerbes}, the relevant geometric structure is that of a gerbe on $M$, and in subsequent sections we attempt to find a suitable modification of geometric quantisation appropriate to this setting. In Sections~\ref{sec:2catBGrb} and~\ref{sec:Bgerbeshigher} we explain in detail the sense in which gerbes may be regarded as categorifications of line bundles, which leads to a better understanding of gerbes and of higher geometry in general. In particular, the discussion of 2-bundle metrics in Section~\ref{sect:2-bdl_metric} refines the current understanding of bundle gerbes as higher line bundles to show that hermitean bundle gerbes can naturally be considered as higher \emph{hermitean} line bundles. The higher analogue of Hilbert spaces of physical states in quantum theory should now be a suitable notion of 2-Hilbert spaces, which we describe in some generality in Section~\ref{sect:2-HSpaces} in the form that we need in this paper. In Section~\ref{sect:2Hspace_of_BGrb} we then construct the higher version of the prequantum Hilbert space of geometric quantisation, and illustrate our approach by working out a number of explicit examples of these 2-Hilbert spaces in Section~\ref{sec:Examples} which are of relevance in closed string theory and M-theory.

In Section~\ref{sec:open} we highlight some of the open questions
which are not yet addressed by our formalism. One of these concerns is an
already notoriously difficult problem in ordinary geometric
quantisation: The second step in that procedure involves the extra
choice of a polarisation, which corresponds to locally representing
the symplectic manifold $M$ as a cotangent bundle $T^*U$. Global
polarisations are not guaranteed to exist, and as yet there is no
general criterion for a symplectic manifold to be quantisable in this sense. In addition, demonstrating that the result of quantisation is independent of all these auxilliary choices on the symplectic manifold $(M,\omega)$ is also a very difficult task (in fact in almost every approach to quantisation). The situation in higher quantisation is even less clear, and we do not deal with these issues in the present paper, but rather focus on prequantisation only.

This article is a companion to the longer paper~\cite{BSS--HGeoQuan} which, amongst other things, developed new categorical structures on morphisms of bundle
gerbes, initiated steps towards higher geometry by introducing
2-bundle metrics, and showed how to obtain a 2-Hilbert space from the
category of sections of a bundle gerbe. The present paper is partly a
summary of the main constructions from~\cite{BSS--HGeoQuan}, presented
in a somewhat more informal way that we hope will be accessible to a
broader readership; we refer to~\cite{BSS--HGeoQuan} throughout for
all technical details. Several points here are elucidated in more
detail, particularly in our discussions of 2-Hilbert spaces in
Sections~\ref{sect:2-HSpaces} and~\ref{sect:2Hspace_of_BGrb}. We have
also worked out some illustrative examples more thoroughly in
Section~\ref{sec:Examples}, obtaining a much sharper statement
concerning dimensional reductions of sections of a bundle gerbe to
sections of a corresponding prequantum line bundle, as is relevant in
reductions of M-theory to string theory: We show that our dimensional
reduction maps descend to 2-isomorphism classes of sections of bundle
gerbes (representatives of which are ``constant'' along the
M-theory direction), and are moreover compatible with the respective monoidal structures and module actions. Our final example in Section~\ref{sect:lens} is new, and it nicely illustrates the construction of a 2-Hilbert space of a non-trivial bundle gerbe with torsion Dixmier-Douady class that is relevant for a certain M-theory dual of closed strings described by an asymmetric cyclic orbifold of the $\sfS\sfU(2)$ WZW model; in particular, this example suggests a general prescription for our higher prequantisation that mimics the well-known situation in ordinary quantisation: Given a higher prequantisation of $M$ and a group $\sfG$ acting freely on $M$, then one obtains a higher prequantisation of the quotient $M/\sfG$ by taking the $\sfG$-invariant part of the higher prequantisation of~$M$.

\subsection*{Glossary of notation}

For the reader's convenience, we summarise here the notational
conventions for categories used throughout this paper:
\begin{myitemize}
	\item $\Hilb$, $\Vect$: \ The category of finite-dimensional complex Hilbert spaces/vector spaces with linear maps.
	\item $\Hilb_\infty$, $\Vect_\infty$: \ The category of possibly infinite-dimensional complex Hilbert spaces/vector spaces.
	\item $2\Hilb$: \ The 2-category of 2-Hilbert spaces.
	\item $\HVBdl(M)$: \ The category of hermitean vector bundles on $M$ with smooth  fibrewise-linear maps.
	\item $\HVBdl^\nabla(M)$: \ The category of hermitean vector bundles with connection on $M$, where the metric is parallel with respect to the connection, with parallel morphisms of vector bundles.
	\item $\HLBdl(M)$, $\HLBdl^\nabla(M)$: \ The subcategory of hermitean line bundles on $M$ without/with connection.
	\item $\BGrb(M)$, $\BGrb^\nabla(M)$: \ The 2-category of hermitean bundle gerbes on $M$ without/with connection.
	\item For any $n$-category $\scC$ and objects $a,b \in \scC$, the $(n{-}1)$-category of morphisms from $a$ to $b$ in $\scC$ is denoted $\scC(a,b)$.
Analogously, given 1-morphisms $f,g: a \rightarrow b$ in $\scC$ we will write $\scC(f,g)$ for the $(n{-}2)$-category of morphisms from $f$ to $g$ in $\scC$.
\end{myitemize}

\section{Geometric quantisation}
\label{sect:GeoQuan_review}

Geometric quantisation starts from a symplectic manifold $(M,\omega)$, i.e. a manifold $M$ of dimension $2n$ with a non-degenerate closed 2-form $\omega$; this is part of the data of most classical physical systems.
Such a manifold is called \emph{prequantisable} if there exists a hermitean line bundle with connection $(L,\nabla^L)$ on $M$ whose magnetic flux is $F^L =  2\pi\, \omega$; by Dirac charge quantisation this is equivalent to the statement that the de Rham class of $\omega$ lies in the image of the map $\rmH^2(M,\RZ) \rightarrow \rmH^2_\dR(M)$.
Sections of $L$ then form the space of wavefunctions for the quantum theory.
For compact $M$ the space $\Gamma(M,L)$ of smooth sections carries a non-degenerate, positive-definite inner product.
If we denote by $h$ the hermitean metric on $L$, it is given by
\begin{equation*}
	(\psi, \phi) \longmapsto \< \psi, \phi \>_{\Gamma(M,L)} = \int_M\, h(\psi, \phi)\, \frac{\omega^n}{n!}\ .
\end{equation*}
The pair $\CH_\pre(L) = (\Gamma(M,L), \<-,-\>_{\Gamma(M,L)})$ then forms a pre-Hilbert space whose Hilbert space completion we denote by $\CH(L) = \rmL^2(M,L)$.\footnote{If $M$ is non-compact we can instead consider the pre-Hilbert space to consist of compactly supported smooth sections.}
It is called the \emph{prequantum Hilbert space} associated with $(M,\omega)$.

In the classical theory, observables are real-valued functions $f \in C^\infty(M,\FR)$.
In the quantum theory they act on the Hilbert space as follows.
First of all, to every such function there is an associated Hamiltonian vector field $X_f$ given by $\iota_{X_f} \omega = - \dd f$, where $\iota_{X}$ denotes insertion of a vector field $X$ into the first slot of a differential form. The vector field
$X_f$ is uniquely defined because of the non-degeneracy of $\omega$.
On wavefunctions $\psi\in\CH_\pre(L)$ we can now set $\CO_f \psi
=-\iu\,\hbar\, \nabla^L_{X_f} \psi + 2\pi\, \hbar\, f\, \psi$.
One then has 
\begin{equation*}
[\CO_f, \CO_g] = -\iu\,\hbar\, \CO_{\{ f,g \}} \ , 
\end{equation*}
where $\{-,-\}$ is the Poisson bracket on $C^\infty(M,\FR)$,
\begin{equation*}
	\{ f,g \} = \iota_{X_f \wedge X_g} \omega\ .
\end{equation*}
The underlying Lie algebra of the Poisson algebra $(C^\infty(M,\FR),
\{-,-\})$ is a central extension of the Lie algebra of Hamiltonian
vector fields; in particular $[X_f,X_g] = - X_{\{f,g\}}$. The
assignment $f\mapsto \CO_f$ of quantum operators to classical functions is called the \emph{Kostant-Souriau prequantisation map}, and continuously extending the action of observables to $\CH(L)$ yields a representation of this Poisson algebra on the prequantum Hilbert space.

As an example, consider a 1-connected manifold $M$ with a symplectic
form $\omega \in \Omega^2(M)$ which has integer periods.\footnote{We
  take an $n$-connected space $M$ to have homotopy groups $\pi_i(M) = 0$ for all $i = 0,1,\ldots,n$.}
Fixing a base-point of $M$ we obtain the diagram
\begin{equation*}
	\xymatrix{
		\CD^2 M \ar@{->}[d]_-{\partial} & \\
		\Omega M \ar@<-0.5ex>[r] \ar@<0.5ex>[r] & \CP M \ar@{->}[d]_-{\partial}\\
		 & M
	}
\end{equation*}
Here $\CP M$ denotes the based path space of $M$,\footnote{Technically we should require based paths with sitting instants here for concatenation of paths to give smooth paths. In the following all mapping spaces are assumed to consist of maps having appropriate sitting instants to make all the necessary gluing of maps smooth.} $\Omega M$ the based loop space, and $\CD^2 M$ denotes the space of based disks $f: D^2 \rightarrow M$ in $M$.\footnote{The base-point of $D^2 \subset \FR^2$ is the point $(1,0)$, such that $S^1 \hookrightarrow D^2$ is a map of pointed manifolds.} 
The map $\CP M \rightarrow M$ is generically a surjective submersion, but for the map $\CD^2 M \rightarrow \Omega M$ to be a surjective submersion we must require 1-connectedness of $M$.
The horizontal arrows restrict a loop to its first and second halves, reversing the orientation of the second half in order to obtain a based map.
Any disk $f: D^2 \rightarrow M$ allows us to integrate the pullback of $\omega$ over $D^2$.
Setting $\lambda(f) = \exp( 2\pi\, \iu\, \int_{D^2}\, f^* \omega)$ defines a map $\lambda: \CD^2M \rightarrow \sfU(1)$.
Two different maps $f, f': D^2 \rightarrow M$ which induce the same map on the boundary yield a map $f \cup_{S^1} f' : S^2 \rightarrow M$, and because $\omega$ has integer periods, $\lambda(f) = \lambda(f'\,)$.
This implies that $\lambda$ descends to a map $\hat{\lambda}: \Omega M \rightarrow \sfU(1)$.
By the same reasoning we see that $\lambda$ defines a cocycle in the sense that for triples $(\gamma_0, \gamma_1, \gamma_2)$ of based paths with common endpoint, we have
\begin{equation}\label{eq:cocyclecond}
	\hat{\lambda} (\overline{\gamma_0} * \gamma_1)\ \hat{\lambda} (\overline{\gamma_1} * \gamma_2) = \hat{\lambda} (\overline{\gamma_0} * \gamma_2)\ ,
\end{equation}
where $*$ denotes concatenation of paths, and $\overline{\gamma_i}$ is the path $\gamma_i$ with opposite orientation.
Thus $\hat{\lambda}$ defines a hermitean line bundle $L$ over $M$, called the \emph{tautological line bundle} of $(M,\omega)$ (see e.g.~\cite{Johnson:2002tc}).
A connection on this line bundle is constructed from $\omega$ via its transgression to $\CP M$, $A_{|\gamma} (X) = -2\pi\, \int_{[0,1]}\, \gamma^*\omega(X,-)$.
Its magnetic flux is given by $F^L = 2\pi\, \omega$, whence this line bundle is a prequantum line bundle for the symplectic manifold $(M,\omega)$.
If $M$ is not 1-connected, no general construction of a prequantum
line bundle is known, and one has to work with the primitive statement
that there exists a line bundle representative for each closed $\omega \in \Omega^2(M)$ with integer periods.

Prequantisation is only the first of two steps in the program of geometric quantisation.
The Hilbert space $\CH(L)$ it produces is usually too big to properly describe the physical system.
The actual physical Hilbert space is a subspace or a quotient of $\CH(L)$.
This redundancy can be understood as follows.
In classical mechanics the symplectic manifold describing the physical system is not the {configuration space} $Q$ (which might be odd-dimensional), but rather it is the {phase space} of the system, which is given by the cotangent bundle $M = T^*Q$ of the configuration space.
This space always has twice the dimension of $Q$ and carries a canonical symplectic structure.
Moreover, all homotopy groups of $T^*Q$ and $Q$ are isomorphic so that $T^*Q$ is 1-connected if and only if $Q$ is 1-connected.\footnote{This follows from the homotopy long exact sequence together with contractibility of the fibres of $T^*Q$.}
As an example consider a particle propagating in $Q = \FR^n$.
Then $T^*Q \cong \FR^n {\times} \FR^n$, and $\omega = \dd p_i \wedge \dd q^i$ where $q^i$ are coordinates for the {position} space $\FR^n$ and $p_i$ are coordinates for the {momentum} space $\FR^n$.
The prequantum line bundle $L$ in this case is trivial, $(L,\nabla^L) = ((\FR^n {\times} \FR^n) {\times} \FC,\, \dd +2\pi\,\iu\, p_i\, \dd q^i)$.
Elements of $\CH(L)$ are then just square-integrable complex-valued functions on $\FR^n {\times} \FR^n$.
However, $\CH(L)$ is not the Hilbert space for the quantum mechanics of a particle on $\FR^n$.
Usually, one chooses either position representation {or} momentum representation since the particle is already completely described by functions depending solely on either the position coordinates or the momentum coordinates.
As sections of $L$, this would mean that these functions are covariantly constant, or parallel, in half the directions of $T^*Q$.
This reduction of the degrees of freedom and the prequantum Hilbert space $\CH(L)$ is called \emph{polarisation}.
It is usually a choice which has to be made by hand, based on physical requirements.
On a generic symplectic manifold $M$, a splitting of $M$ into a product does not exist, but locally a splitting of the directions on $M$ into two halves is still possible by Darboux's theorem.
In this case polarisation amounts to choosing a foliation of $M$ which is Lagrangian for $\omega$, and then considering sections that are parallel along the foliation.
Other methods of polarisation exist as well, see e.g.~\cite{SS--Groupoids_loops_GeoQuan,SS--Quan_of_2plectic_mfds,Rogers--Thesis} and references therein.

\section{Strings, membranes and fluxes}
\label{sect:strings_and_tautological_gerbe}

The main physical motivation for the present paper comes from the
problem of quantising closed strings and open M2-branes in the
backgrounds of certain fluxes. The situation described in
Section~\ref{sect:GeoQuan_review} has an analog for open strings in a
$B$-field background, wherein the boundaries of strings ending on a
D3-brane quantise the D3-brane worldvolume
$M$~\cite{Seiberg-Witten--Strings_and_NCG,Szabo:2001kg}; the closed Kalb-Ramond
field $B\in \Omega^2(M)$ serves as a magnetic flux on $M$ to
which the techniques of standard geometric quantisation are
applicable. The situation changes, however, when we lift this
configuration to M-theory via T-duality, where it describes open
M2-branes ending on an M5-brane with a constant 3-form $C$-field. In
this case the geometry of the M5-brane worldvolume is
described by replacing the Poisson bracket with a 3-bracket 
related to the $C$-field, see
e.g.~\cite{Chu:2009iv,SS--Quan_of_2plectic_mfds,Saemann:2012ex} and
references therein. The changes incurred in the M-theory lift are the
replacements of the 2-form $B$-field with the 3-form $C$-field and of
the particle-like boundaries of open strings with the closed
string-like boundaries of open membranes, so that the noncommutative
geometry arising from quantising the Poisson bracket is replaced by a
nonassociative geometry quantising a 3-bracket. This kind of situation in fact
arises directly in closed string theory with a supergravity
background: Simplifying the system by dropping the fermions, the
dilaton and the metric, as well as higher form fields, this leaves us with a closed string propagating through a target space $M$ in the presence of an NS--NS $H$-flux, which is described by a closed 3-form $H \in \Omega^3(M)$ such that $H/2\pi$ has integer periods by virtue of (generalised) Dirac charge quantisation. We shall write $H=2\pi\,\varpi$ where $\varpi$ represents a class in $\rmH^3(M,\RZ)$.

The immediate observation at this point is that {geometric quantisation does not apply to these systems}, simply because we are not in the symplectic setting.
We therefore seek a modification of geometric quantisation which can treat situations like this.

First of all, the notion of non-degeneracy straightforwardly carries over to 3-forms~\cite{Baez:2008bu,Rogers--Thesis}:
We call $\varpi$ non-degenerate if $\iota_X \varpi = \varpi(X,-,-)$ is zero if and only if the vector field $X$ is zero.
A closed non-degenerate 3-form $\varpi$ is called a \emph{2-plectic form}, and the pair $(M,\varpi)$ is called a \emph{2-plectic manifold}.
Thus we have found a modification of the notion of a symplectic
structure, which was the starting point in geometric quantisation.

Prequantisation then relied upon a Chern-Weil realisation of $\omega \in \Omega^2(M)$ in terms of a hermitean line bundle with connection.
Again for the simple reason that we are now dealing with a 3-form, such a geometric model can no longer be given by a line bundle on $M$.
One way out of this is to say that the actual configuration space of a closed string should be the free loop space of $M$, i.e. $LM = C^\infty(S^1,M)$, or the space of unparameterised oriented loops $\CL M =C^\infty(S^1,M)/ \Diff_+(S^1)$, where $\Diff_+$ denotes the orientation-preserving diffeomorphisms; we could then attempt to apply the techniques of Section~\ref{sect:GeoQuan_review} on loop space.
However, on the first space the 2-form induced by $\varpi$ via
transgression is degenerate (insert a vector field tangent to the
loop), and on both spaces it is extremely difficult to find good
notions of square-integrability: Since these spaces are not locally compact the notion of compactly supported sections does not seem appropriate, as heuristically such sections would have support on sets we expect to have measure zero.
From a more fundamental perspective, string theory is a quantum theory of gravity and so is expected to quantise spacetime itself rather than its loop space.
Certain features of a quantisation of $M$ will also translate to
quantum objects on its loop space, but the fundamental quantisation
happens in the target space itself.
The best example of this is in non-geometric string theory where noncommutative and nonassociative geometries of spacetimes arise as T-duals of geometric backgrounds carrying 3-form $H$-fluxes (see for instance~\cite{Blumenhagen:2010hj,Lust:2010iy,BDL--NonAsso,Condeescu:2012sp,Mylonas:2012pg,Aschieri-Szabo--NonAsso}). 

We therefore take a different route.
Recall that in Section~\ref{sect:GeoQuan_review} we were able to give
a construction of a geometric representative of a symplectic form
$\omega$ with integer periods on a 1-connected manifold, taking only that form as input.
The key observation there was that we can integrate the 2-form over disks, and that on 2-spheres this produces an integer.
We can try to lift this construction to 3-forms and analyse the structure we obtain.
In the case of a 2-plectic form $\varpi$ with integer periods we could integrate
over balls; analogously, on 3-spheres we then obtain integer values.
Assuming this time that $M$ is 2-connected, we thus obtain the following diagram in which vertical arrows are surjective submersions:

\begin{equation}
\label{diag:taut_gerbe}
	\xymatrix{
		\CD^3 M \cong \CP \Omega^2 M \ar@{->}[d]_-{\partial} & & \\
		 \CS^2 M \cong \Omega^2 M \ar@<-0.5ex>[r] \ar@<0.5ex>[r] & \CD^2 M \cong \CP \Omega M \ar@{->}[d]_-{\partial} & \\
		 & \Omega M \ar@<-0.5ex>[r] \ar@<0.5ex>[r] & \CP M \ar@{->}[d]_-{\partial}\\
		 & & M
	}
\end{equation}
In each step of this stair diagram one just adds another path $\CP$
and loop $\Omega$ iteration:
Since the mapping spaces are based, there are identifications $\Omega^n M \cong \CS^n M = C^\infty_*(S^n,M)$ and $\CP \Omega^n M \cong \CD^{n+1} M = C^\infty_*(D^{n+1},M)$, where the subscript $*$ indicates that {based} smooth maps are used.

Using integration of pullbacks of $\varpi$ over $D^3$, we now obtain a function $\sigma: \CD^3 M \rightarrow \sfU(1)$ by setting $ \sigma(f) = \exp( 2\pi\, \iu\, \int_{D^3}\, f^*\varpi)$ for maps $f: D^3 \rightarrow M$.
On two balls with coinciding boundary the values of $\sigma$ agree because of the integrality of $\varpi$.
Hence $\sigma$ descends to a $\sfU(1)$-valued function $\hat{\sigma}$ on $\CS^2 M$.
As in Section~\ref{sect:GeoQuan_review}, $\hat{\sigma}$ satisfies a
cocycle condition on 2-spheres with one coincident hemisphere, and thus defines a hermitean line bundle $L$ over $\Omega M$.
A connection on this hermitean line bundle is, again as in the
construction of Section~\ref{sect:GeoQuan_review}, given by the
transgression of $\varpi$ to $\CD^2 M$, $A_{|f} = -2\pi\, \int_{D^2}\, f^*\varpi$.
The magnetic flux of this connection is given by the transgression of $\varpi$ to $\Omega M$,
\begin{equation*}
	F^L_{|\gamma} = 2\pi\, \int_{S^1}\, \gamma^*\varpi\ .
\end{equation*}

We now observe that the transgression of any differential form to a loop $\gamma = \overline{\gamma_0} * \gamma_1$ naturally splits into the difference of its transgressions to the paths $\gamma_0$ and $\gamma_1$.
In our situation we get
\begin{equation*}
	F^L_{|\gamma} = 2\pi\, \int_{S^1}\, (\overline{\gamma_0} * \gamma_1)^* \varpi
	= 2\pi\, \int_{[0,1]}\, \gamma_1^* \varpi - 2\pi\, \int_{[0,1]}\, \gamma_0^* \varpi
	= B_{|\gamma_1} - B_{|\gamma_0}\ ,
\end{equation*}
with the 2-form $B \in \Omega^2(\CP M)$ chosen as $B_{|\gamma} = 
2\pi\, \int_{[0,1]}\, \gamma^*\varpi$.%
\footnote{The vector fields that forms of this type on mapping spaces act on are inserted into the first slots of the form under the integral.}
Finally, recall that the magnetic flux of the line bundle we
constructed from a 2-form in Section~\ref{sect:GeoQuan_review} was
related to that 2-form.
Here too we are brought full circle by observing that $\dd B = 2\pi\,  \partial^*\varpi$, i.e. $\dd B$ descends to $M$ giving precisely $ 2\pi\,  \varpi=H$.
Thus we say that we have represented the 3-form $\varpi$ in a Chern-Weil manner, and we call $(M,\varpi)$ \emph{prequantisable}.

There is yet an additional observation we can make about this construction.
The fibre of $L$ over a loop $\gamma$ is given by equivalence classes of pairs $(f, z)$, where $z \in \FC$ and $f: D^2 \rightarrow M$ is a based disk in $M$ whose boundary is $\gamma = \partial f$.
Two such pairs $(f,z)$ and $(f',z'\,)$ are equivalent if $z' = \exp( 2\pi\, \iu\, \int_{D^3}\, h^*\varpi)\, z$, where $h: D^3 \rightarrow M$ is any ball which is bounded by the 2-sphere defined by gluing $\overline{f}$ and $f'$.
Consider again three based paths $\gamma_i$, $i=0,1,2$ in $M$ with common endpoint.
Choose three disks $f_{ij}: D^2 \rightarrow M$ such that $\partial f_{ij} = \overline{\gamma_i} * \gamma_{j}$ for $i<j$, i.e. the boundary of the disk $f_{ij}$ in $M$ is the loop defined by $\gamma_i$ and $\gamma_j$.
We now get two disks bounded by $\overline{\gamma_0} * \gamma_2$, namely $f_{02}$ and $f_{01} \cup_{\gamma_1} f_{12}$.
Since $\pi_2(M) = 0$ there exists a ball $g: D^3 \rightarrow M$ whose boundary is composed of precisely those disks glued along their common bounding loop.
We can map given elements $[f_{01},\, z_{01}] \in L_{\overline{\gamma_0} * \gamma_1}$ and $[f_{12},\, z_{12}] \in L_{\overline{\gamma_1} * \gamma_2}$ as
\begin{equation*}
	 L_{\overline{\gamma_0} * \gamma_1} \otimes
         L_{\overline{\gamma_1} * \gamma_2} \ \ni \ [f_{01},\, z_{01}] \otimes [f_{12},\, z_{12}]
	 \longmapsto \big[ f_{01} \cup_{\gamma_1} f_{12},\, z_{01}\,
         z_{12} \big] \ \in \ L_{\overline{\gamma_0} * \gamma_2}\ .
\end{equation*}
We can further modify this map by using the equivalence relation in
the definition of the fibres of $L$ to get
\begin{equation*}
	\big[ f_{01} \cup_{\gamma_1} f_{12},\, z_{01}\, z_{12} \big] =
        \Big[ f_{02},\, \exp \Big( 2\pi\, \iu\, \int_{D^3}\, g^*\varpi
        \Big)\, z_{01}\, z_{12} \Big]\ .
\end{equation*}
This yields an isomorphism
\begin{equation}
\label{eq:taut_mu}
	\mu_{\gamma_0,\gamma_1,\gamma_2}: L_{\overline{\gamma_0} *
          \gamma_1} \otimes  L_{\overline{\gamma_1} * \gamma_2}
        \arisom L_{\overline{\gamma_0} * \gamma_2}\ .
\end{equation}
This structure is analogous to the cocycle condition \eqref{eq:cocyclecond}, and is usually referred to as a multiplication.
The isomorphism \eqref{eq:taut_mu} is compatible with the connection we found above.

Thus we obtain a geometric structure which realises a closed 3-form $\varpi$ on $M$ with integer periods, provided that $M$ is 2-connected.
This is analogous to how the tautological line bundle in
Section~\ref{sect:GeoQuan_review} realised a closed 2-form with
integer periods on a 1-connected manifold.
The geometric structure we have unveiled here is called the \emph{tautological bundle gerbe} of $(M,\varpi)$, and appeared originally in~\cite{Murray--BGerbes}.
In Section~\ref{sect:Bundle_gerbes} we will abstract the structures present in this example, thereby arriving at the general notion of a \emph{bundle gerbe}.

\section{Bundle gerbes and $B$-fields}
\label{sect:Bundle_gerbes}

In the construction of the tautological bundle gerbe it was crucial
that $M$ be 2-connected in order to ensure that the vertical maps in
the diagram~\eqref{diag:taut_gerbe} are surjective.
For a general bundle gerbe, we therefore start with a surjective submersion $\pi: Y \rightarrow M$.\footnote{The submersion property ensures that the fibres are smooth manifolds.}
We then get the two horizontal maps $p_i: Y^{[2]} \rightrightarrows Y$
automatically, where $p_i$ is defined to {forget} the $i$-th entry in
$Y^{[2]} = \{ (y_0,y_1) \in Y^2\, |\, \pi(y_0) = \pi(y_1)
\}$.\footnote{This is the convention that comes from viewing $Y^{[2]} \rightrightarrows Y$ as part of a simplicial set (the nerve of the \v{C}ech groupoid of $Y \rightarrow M$) and taking $p_i$ to be the respective face maps.}
The second layer in the diagram~\eqref{diag:taut_gerbe} is just the tautological line bundle on $\Omega M$.
We formalise this by taking a hermitean line bundle $L \rightarrow Y^{[2]}$.
Recall the multiplication structure on $L$ from~\eqref{eq:taut_mu}.
This is made part of the data by demanding the existence of a multiplication isomorphism of hermitean line bundles $\mu: p_2^* L \otimes p_0^* L \arisom p_1^* L$ over $Y^{[3]}$, which is associative over $Y^{[4]}$.
The 3-form on $M$ was then reconstructed from a connection on $L$ which was compatible with $\mu$.
Hence we demand that $L$ be endowed with a connection $\nabla^L$ for which $\mu$ is parallel with respect to the connections induced on its source and target line bundles.
This implies that its magnetic flux $F^L \in \Omega^2\big(Y^{[2]}\big)$ satisfies $p_2^*F^L + p_0^* F^L = p_1^*F^L$ in $ \Omega^2\big(Y^{[3]}\big)$.
In turn, this allows us to find a 2-form $B \in \Omega^2(Y)$ such that
$p_0^*B - p_1^*B = F^L$, just as we were able to write down explicitly
for the tautological bundle gerbe; this follows from exactness of the
\v{C}ech complex $\check{C}^\bullet(\pi: Y \rightarrow M,\, \Omega^k)$
of sheaves of $k$-forms for each $k$ (see~\cite{Murray--BGerbes}).
There is some ambiguity in picking such a 2-form as the relation between $B$ and $F^L$ only determines $B$ up to \v{C}ech-exact forms.
A choice of such a 2-form is called a \emph{curving} of the bundle gerbe, and a \emph{connection on a bundle gerbe} or a \emph{$B$-field} consists of a connection on $L$ together with a curving.
Finally, we see that $p_1^*\, \dd B - p_0^*\, \dd B = \dd F^L = 0$ by the Bianchi identity for $\nabla^L$.
Again by exactness of the \v{C}ech complex of $k$-forms we therefore obtain a 3-form $H$ on $M$ such that $\pi^* H = \dd B$.
In this case $H$ is uniquely defined once a curving $B$ has been chosen, since $\pi^*$ is injective.
The 3-form $H$ is called the \emph{$H$-flux of the bundle gerbe}.
To summarise, a {hermitean bundle gerbe with connection} is given by data $\CL = (L, \mu, Y, \nabla^L, B)$.
In the following the terminology bundle gerbe will always refer to a hermitean bundle gerbe.

We have already seen an example of a bundle gerbe in Section~\ref{sect:strings_and_tautological_gerbe}.
Let us here consider a surjective submersion over $M$ given by taking $Y = \CU =
\bigsqcup_{i \in \Lambda} \, U_i$ to be the total space of a good open
covering of $M$.\footnote{An open covering $\{U_i\}_{i\in\Lambda}$ is \emph{good} if all $U_i$ and all possible finite intersections are contractible.}
In this case, $Y^{[2]} = \CU^{[2]}= \bigsqcup_{i,j \in \Lambda} \, U_{ij}$, where $U_{ij} = U_i \cap U_j$, and a bundle gerbe defined using a surjective submersion of this kind is called a \emph{local bundle gerbe}.
A hermitean line bundle over $\CU^{[2]}$ consists of a hermitean line bundle $L_{ij} \rightarrow U_{ij}$ for every $(i,j) \in \Lambda^2$, and as the covering is good we can choose these to be trivial as hermitean line bundles {without connection}.
The only non-trivial input for a bundle gerbe without connection and surjective submersion given by the total space of a good cover is hence the isomorphism $\mu$.
This is now an isomorphism of trivial line bundles $\mu_{ijk}: L_{ij}
\otimes L_{jk} \arisom L_{ik}$ and hence corresponds to a collection
of functions $\mu_{ijk}: U_{ijk} \rightarrow \sfU(1)$ on triple intersections.
The associativity constraint on $\mu$ translates to
\begin{equation*}
	\mu_{ikl}\, \mu_{ijk} = \mu_{ijl} \, \mu_{jkl} \ ,
\end{equation*}
for all $i,j,k,l \in \Lambda$. Thus $(\mu_{ijk})$ defines a $\sfU(1)$-valued \v{C}ech 2-cocycle on $M$.
Enriching this data by the components of a connection on the bundle gerbe, one is left with a triple $(\mu_{ijk}, A_{ij}, B_i)$, where $A_{ij}$ is the connection 1-form of the connection on $L_{ij}$ and $B_i = B_{|U_i}$.
A triple such as this, obtained from a bundle gerbe defined over $Y =
\CU$, defines a \v{C}ech representative of a Deligne 2-cocycle in
degree~2 on $M$; we denote the corresponding class of the gerbe $\CL$
in Deligne cohomology by $\rmD(\CL) \in \rmH^2(M,\CD^\bullet(2))$.
Forgetting the data of differential forms in the Deligne cocycle we recover the \v{C}ech
$\sfU(1)$-cocycle $(\mu_{ijk})$; its image in integer cohomology is called the \emph{Dixmier-Douady class} of $\CL$ and is denoted by $\DD(\CL) \in \rmH^3(M,\RZ) \cong \rmH^2(M,\sfU(1))$.

A bundle gerbe which we will frequently encounter in the ensuing sections is the \emph{trivial bundle gerbe}.
It is defined by the data $\CI = (M {\times} \FC,\, \cdot\, , M)$.
Its covering of $M$ is the identity covering $1_M: M \rightarrow M$, so that $M^{[2]}$ identifies with $M$.
The hermitean line bundle over this space is the trivial hermitean line bundle $I = M {\times} \FC \rightarrow M$.
We introduce a bundle gerbe multiplication using the usual multiplication on $\FC$, i.e. $(z, z'\,) \mapsto z \cdot z'$.
Additionally, just as the trivial line bundle can carry a non-trivial
connection given by a 1-form on the base, the trivial bundle gerbe can
still carry a non-trivial $B$-field.
We define $\CI_\rho = (M {\times} \FC,\, \cdot\, ,M, \dd, \rho)$, where $\dd$ denotes the trivial connection on the trivial line bundle, and $\rho \in \Omega^2(M)$ is an arbitrary 2-form defining the curving of $\CI_\rho$.
The $H$-flux of $\CI_\rho$ is given by $H= \dd \rho \in \Omega^3(M)$.

Another interesting example of a bundle gerbe occurs in the form of the tautological gerbe constructed in Section~\ref{sect:strings_and_tautological_gerbe}.
Any compact, simple, simply connected Lie group $\sfG$ with Lie algebra $\frg$ is automatically 2-connected and has $\rmH^3(\sfG,\RZ) \cong \RZ$ with generator given by the de~Rham class of $\varpi = \frac{1}{3!}\, \<\mu_\sfG,[\mu_\sfG,\mu_\sfG]\>_{\frg}$, where $\mu_\sfG \in \Omega^1(\sfG,\frg)$ denotes the Maurer-Cartan form of $\sfG$~\cite{PS--Loop_groups}.
Thus we can apply the tautological bundle gerbe construction to every pair
$(\sfG, k\, \varpi)$ for $k \in \RZ$.
The Lie group operations are smooth, whence we obtain group structures
on the relevant spaces of smooth maps into $\sfG$ and also on the line bundle $L \rightarrow \Omega \sfG$.
We construct in this way the central extensions $\widehat{\Omega_k\sfG} = L$ of the loop group $\Omega \sfG$, together with the additional data of a connection on the extension bundle and a curving 2-form on $\CP \sfG$.

\section{The 2-category of bundle gerbes}
\label{sec:2catBGrb}

In this section we start by summarising the 2-categorical theory of bundle gerbes as initiated in~\cite{MS--BGrbs-stable_isomp_and_local_theory} and generalised in~\cite{Waldorf--More_morphisms}.
The material in Sections~\ref{sect:direct_sum} and~\ref{sect:Structure_on_2-morphisms} is new.
Details and proofs can be found in~\cite{BSS--HGeoQuan}.

\subsection{Local picture}
\label{sect:BGrb_Mors_loc}

The example of the local bundle gerbe from Section~\ref{sect:Bundle_gerbes} has an illuminating intuitive interpretation.
The line bundles $L_{ij} \rightarrow U_{ij}$ satisfy
\begin{equation*}
	L_{ij} \otimes L_{jk} \cong L_{ik} \ ,
\end{equation*}
for all $i,j,k\in\Lambda$. We can compare this to the transition data of a hermitean line bundle over the same open cover, which is given by transition functions $g_{ij} : U_{ij} \rightarrow \sfU(1)$ subject to the cocycle condition
\begin{equation*}
 g_{ij}\, g_{jk} = g_{ik} \ ,
\end{equation*}
for all $i,j,k\in\Lambda$.

Hence the heuristic behind passing from hermitean line bundles to hermitean bundle gerbes is that $\sfU(1)$-valued transition functions get replaced by $\invs{\Hilb}$-valued transition functions.
Here $\Hilb$ denotes the category of finite-dimensional Hilbert spaces and linear maps.
Among other structures, it carries a tensor product $\otimes: \Hilb \times \Hilb \rightarrow \Hilb$.
This is \emph{symmetric} in the sense that for every $V,\, V' \in \Hilb$ there exists a distinguished isomorphism $V \otimes V' \cong V' \otimes V$ which satisfies certain coherence conditions.
The \emph{unit object} with respect to the tensor product is $\FC$ with its natural inner product.
A Hilbert space $V \in \Hilb$ is called \emph{invertible} if there is another Hilbert space $V' \in \Hilb$ and an isomorphism $V \otimes V' \cong \FC$.
In this case we call $V'$ a \emph{weak inverse} for $V$.
We denote the subcategory of invertible objects and unitary morphisms by $\invs{\Hilb}$.
For every Hilbert space $V$ there exists a \emph{dual} Hilbert space $V^*$, and it is related to $V$ via the Riesz isomorphism
\begin{equation*}
	\theta: V \arisom \overline{V^*} \ , \quad \psi \longmapsto \theta(\psi) = \< \psi, - \>_V\ .
\end{equation*}
If $V$ is invertible, then its inverse is represented by $V^*$, which
is isomorphic to the complex conjugate $\overline{V}$.
In this sense the invertible objects in $\Hilb$ are also \emph{unitary}; they are precisely the 1-dimensional Hilbert spaces.
The category $\invs{\Hilb}$ is closed under the tensor product, which is symmetric, and each object has a weak inverse.
By construction, all morphisms in this category are invertible.
We call a category with these properties a \emph{2-group}.%
\footnote{There are two versions of a categorified group, namely a
  category in groups, or a group in categories.
The category $(\invs{\Hilb},\otimes)$ is a symmetric monoidal groupoid, i.e. it belongs to the latter class.
We will encounter a similar ambiguity in Section~\ref{sect:2-HSpaces}.}

A function
$L: Y^{[2]} \rightarrow \invs{\Hilb}$ assigns to each point $(y_0,y_1) \in Y^{[2]}$ a 1-dimensional Hilbert space $L_{(y_0,y_1)}$, and we regard such a function as smooth if the resulting family of Hilbert spaces forms a hermitean line bundle over $Y^{[2]}$.
Since in $\invs{\Hilb}$ we now have isomorphisms between objects, the cocycle condition on the transition functions gets weakened from equality to the existence of an isomorphism satisfying a suitable coherence condition (which is precisely the associativity condition on $\mu$).
For connections on the line bundles, the picture is a little less clear; here we shall endow the transition line bundles with connections in a subsequent step.%
\footnote{It would be more precise to say that the {sheaf} of
  $\sfU(1)$-valued transition functions gets replaced by the
  {stack} of transition line bundles with
  connection, and to consider local sections of this stack
  instead of generalised transition functions. We
  refrain from adapting this point of view for pedagogical reasons.}

Following the viewpoint that $L_{ij}$ serve as transition functions of a \emph{higher line bundle}, we can explore what morphisms of this object look like.
Recall that for two ordinary hermitean line bundles with transition functions $g_{ij}$ and $ g'_{ij}$, morphisms are given by collections of maps $e_i: U_i \rightarrow \FC$ which satisfy
\begin{equation}\label{eq:HLBdlmorphism}
	g_{ij}\, e_j = e_i\, g'_{ij}\ ,
\end{equation}
so that these functions can be {glued} along the transition functions $g_{ij}$ and $g'_{ij}$.

The local representatives $e_i$ are functions into the ring $\FC$, which contains the group that the transition functions are valued in.
Translating this to the framework above leads us to consider smooth $\Hilb$-valued functions $E_i: U_i \rightarrow \Hilb$, or in more familiar terms, hermitean vector bundles $E_i \rightarrow U_i$.
Heuristically, we replace the target ring $(\FC,+,\cdot\,)$ by
$(\Hilb,\oplus,\otimes)$; we shall say more about this analogy in Section~\ref{sect:2-HSpaces}.
The gluing relation is weakened to the existence of isomorphisms
\begin{equation}
\label{eq:local_mor}
	\alpha_{ij}: L_{ij} \otimes E_j \arisom E_i \otimes L'_{ij}\ ,
\end{equation}
satisfying a compatibility condition with $\mu$: We can either multiply two of the transition line bundles successively with $E_i$ and then multiply the transition line bundles on the target side, or we can take their product first and then multiply the result with $E_i$, yielding
\begin{equation*}
	(1_{E_i} \otimes \mu'_{ijk}) \circ (\alpha_{ij} \otimes 1_{L'_{jk}}) \circ (1_{L_{ij}} \otimes \alpha_{jk}) = \alpha_{ik} \circ (\mu_{ijk} \otimes 1_{E_k})\ .
\end{equation*}
We can then require that $E_i$ comes endowed with a hermitean connection such that all morphisms above are parallel with respect to the induced connections.

A crucial difference between the line bundle and bundle gerbe cases, which we have already encountered above, becomes fully visible here.
Recall that the cocycle condition and the gluing condition for
morphisms are only required to hold up to coherent isomorphism, i.e. since the data of a morphism of bundle gerbes is based on a hermitean vector bundle with connection, we can still have another level of morphisms between pairs of such data.
Alternatively, observe that the target $\Hilb$ of the local representatives of the sections is a category rather than a set (such as $\FC$), so that there exist non-trivial relations, or morphisms, between the objects.
We call the morphisms on this higher, second level \emph{2-morphisms} of bundle gerbes.
A 2-morphism from $(E_i, \nabla^{E_i}, \alpha_{ij})$ to $(E'_i,
\nabla^{E'_i}, \alpha'_{ij})$ is a collection of morphisms $\psi_i:
E_i \rightarrow E'_i$ of hermitean vector bundles with connection
which we require to be bicovariantly constant.
They are further required to be compatible with the isomorphisms in the sense that
\begin{equation*}
	\alpha'_{ij} \circ (1_{L_{ij}} \otimes \psi_j) = (\psi_i \otimes 1_{L'_{ij}}) \circ \alpha_{ij}\ .
\end{equation*}
In the following we will often make no explicit mention of the connection in the data of a 1-morphism and just write $(E,\alpha)$.

The data for a local bundle gerbe simplifies considerably for the case
of a good open covering because in that situation we can choose all
hermitean line bundles to be topologically trivial, and all the information about
the bundle gerbe is contained in the multiplication isomorphism
$\mu_{ijk}$ and the local data $A_{ij},B_i$ for the $B$-field.
Similarly, in this situation we can always assume (up to 2-isomorphism) that the hermitean vector bundles that make up the morphisms of bundle gerbes are trivial over $U_i$.
Thus the data which is left is the isomorphism $\alpha_{ij}$, which is now a map $\alpha_{ij}: U_{ij} \rightarrow \sfU(n)$, and 1-forms $a_i \in \Omega^1(U_i,\mathfrak{u}(n))$.
The compatibility with $\mu$ and $\mu'$ becomes a \emph{twisted cocycle condition}
\begin{equation}\label{eq:twistedcocycle}
	\mu'_{ijk}\, \alpha_{ij}\, \alpha_{jk} = \alpha_{ik}\, \mu_{ijk}\ .
\end{equation}
For equal source and target bundle gerbe, the two twists $\mu_{ijk}$ and $\mu'_{ijk}$ in this equation cancel, and we are left with transition data for a hermitean vector bundle with connection on $M$;
2-morphisms between such morphisms then precisely provide local representations of parallel morphisms of these hermitean vector bundles with connection. On the other hand, if the target is the trivial bundle gerbe so that $\mu_{ijk}'=1$ in \eqref{eq:twistedcocycle}, then $(\alpha_{ij},\mu_{ijk})$ are the data of a \emph{rank~$n$ twisted vector bundle}~\cite{Karoub--Twisted_bundles_and_twisted_K-theory}, also known as a \emph{bundle gerbe module of rank~$n$}~\cite{BCMMS--Twisted_K-theory}.

\subsection{Global picture}
\label{sect:Rmks_on_global_picture}

In Section~\ref{sect:BGrb_Mors_loc} we introduced morphisms of bundle gerbes which are defined over a common open covering of $M$.
It is possible to generalise this definition of morphisms both to coverings given by general surjective submersions, and to gerbes defined over different coverings, see~\cite{Waldorf--More_morphisms} for details.
In this way, we assemble bundle gerbes on $M$ into a {2-category} $\BGrb^\nabla(M)$.

The observation at the end of  Section~\ref{sect:BGrb_Mors_loc}, that
endomorphisms of local bundle gerbes are related to descent data for hermitean vector bundles with connection on $M$, holds as well in this 2-category.
In particular, there is an equivalence of categories
\begin{equation}
\label{eq:Reduction_functor}
	\rmR: \BGrb^\nabla(M)(\CI_0,\CI_0) \arisom \HVBdl^\nabla(M)\ .
\end{equation}
Here, for $\CL, \CL' \in \BGrb^\nabla(M)$, we denote by $\BGrb^\nabla(M)(\CL, \CL'\,)$ the category of morphisms from $\CL$ to $\CL'$, and $\HVBdl^\nabla(M)$ is the category of hermitean vector bundles with connection on $M$ and morphisms given by parallel, smooth, fibrewise-linear maps.
This equivalence follows from the fact that $\HVBdl^\nabla$ is a {stack}.

As an example of 1-morphisms, consider the tautological bundle gerbe $\CL_k$ over a compact simply connected Lie group $\sfG$ from Section~\ref{sect:Bundle_gerbes}.
Recall that these bundle gerbes are the same as the central extensions of the loop group $\widehat{\Omega_k \sfG} \rightarrow \Omega \sfG$.
Consider a representation $E$ of $\widehat{\Omega_k \sfG}$, which is given by a map $\widehat{\Omega_k \sfG} \otimes E \rightarrow E$. Since the extension is central, the fibre $\sfU(1)$ acts as $\lambda \mapsto \lambda\, 1_E$.
Thus consider the trivial bundle $ \Omega \sfG \times E \rightarrow \Omega \sfG$.
Transition morphisms are given by
\begin{equation*}
	\alpha: \widehat{\Omega_k \sfG} \otimes E \longrightarrow E \ , \quad (\hat{a}, e)_{|b} \longmapsto (\hat{a} e)_{|a\,b}\ ,
\end{equation*}
for a lift $\hat{a} \in \widehat{\Omega_k \sfG}$ of $a \in \Omega \sfG$, and $b \in \Omega \sfG$.
This is of the form of a 1-morphism $\CL_k \rightarrow \CI$ of bundle gerbes without connection.
However, here $E$ has infinite-rank, and bundles with infinite-dimensional fibres are not sufficiently well-behaved for an interesting theory of bundle gerbes,%
\footnote{For example, the trivial bundle gerbe has only a single infinite-rank 1-endomorphism up to 2-isomorphism due to Kuiper's Theorem.
This trivialises the category of endomorphisms, analogously to how adding infinite-rank Hilbert bundles to the category of hermitean vector bundles on $M$ trivialises $\HVBdl(M)$.}
so we restrict ourselves to finite-rank hermitean vector bundles as employed above.

A 1-morphism of bundle gerbes is (weakly) invertible if and only if its underlying hermitean vector bundle is of rank~1.
This stems from the analogous property of objects in $(\Hilb,\otimes)$, which are invertible with respect to $\otimes$ if and only if they are 1-dimensional.
We have seen in Section~\ref{sect:Bundle_gerbes} that every {local} bundle gerbe $\CL$ defines a Deligne class $\rmD(\CL) \in \rmH^2(M,\CD^\bullet(2))$.
With the generalised notion of 1-morphisms of bundle gerbes, it
follows that every bundle gerbe is isomorphic to a local bundle gerbe.
By choosing such an isomorphism and taking the Deligne class we thus obtain a Deligne class for every bundle gerbe; it is independent of the choice of isomorphism of the bundle gerbe with a local bundle gerbe.
From the \v{C}ech representation of Deligne cocycles over a good open
covering, it is  evident that every element of $\rmH^2(M,\CD^\bullet(2))$ arises as the Deligne class $\rmD(\CL)$ of a bundle gerbe $\CL$.
Thus we obtain a well-defined map
\begin{equation*}
	\rmD: \BGrb^\nabla(M) \longrightarrow \rmH^2(M,\CD^\bullet(2))\ ,
\end{equation*}
which descends to a group isomorphism on 1-isomorphism classes of bundle gerbes on $M$.

The construction of bundle gerbes can be rephrased in terms of functors on higher \v{C}ech groupoids.
For every hermitean line bundle $L$, there exists a covering $Y \rightarrow M$ such that $L$ descends from the trivial line bundle over $Y$; this descent data is
the same as a functor $g: \big(Y^{[2]} \rightrightarrows Y\big)
\rightarrow (\FC \rightrightarrows *)$, where composition in the
target category is given by multiplication, while in the source
category (the \v{C}ech groupoid of $Y\to M$) a pair $(y_0,y_1)$ is understood as an isomorphism from $y_0$ to $y_1$.
Morphisms of hermitean line bundles are then natural transformations of such functors.
Again replacing $\FC$ by $\Hilb$, we can now look for functors $L$ into $\invs{\Hilb}$ that define transition data for a bundle gerbe.
Since the target is now the 2-category $\Hilb \rightrightarrows *$, we must replace the \v{C}ech groupoid by a 2-category as well.
We can take this to be the \v{C}ech 2-groupoid $\disc\big(Y^{[2]}
\rightrightarrows Y\big)$, which is the discrete 2-category built from
the \v{C}ech groupoid $Y^{[2]} \rightrightarrows Y$ by adding an identity 2-morphism for each morphism.
Thus a bundle gerbe can be regarded as a 2-functor
\begin{equation*}
	(L,\mu): \disc\big(Y^{[2]} \rightrightarrows Y\big) \longrightarrow (\Hilb \rightrightarrows *)\ .
\end{equation*}
The bundle gerbe multiplication $\mu$ arises as the part of the data of a 2-functor which establishes its weak form of compatibility with composition.
Analogously to the case of line bundles, 1-morphisms and 2-morphisms
arise as natural transformations of functors of 2-categories valued in
$\Hilb$, namely $(E,\alpha): (L,\mu) \Rightarrow (L',\mu'\, )$.

\subsection{Pullbacks, products and duals}
\label{sect:pullproddual}

Bundle gerbes are constructed from local transition or descent data.
As we know from vector bundles, these data can be pulled back along smooth maps.
Accordingly, any smooth map $f: N \rightarrow M$ induces a 2-functor $f^*: \BGrb^\nabla(M) \rightarrow \BGrb^\nabla(N)$, and the assignment $f \mapsto f^*$ is contravariantly functorial.\footnote{In particular, $\BGrb^\nabla$ is a 2-stack~\cite{Nikolaus-Schweigert--Equivar}.}

Transition functions for hermitean line bundles are valued in $\sfU(1)$.
Let $Y = \CU$ be a good open covering of $M$.
Given two such transition functions $g, h: \CU^{[2]} \rightarrow \sfU(1)$ over the same good covering, their pointwise product $g \cdot h: \CU^{[2]} \rightarrow \sfU(1)$ also represents a hermitean line bundle over $M$.
As every hermitean line bundle is isomorphic to one described by transition data over $\CU$, this allows us to define a product on hermitean line bundles over $M$, namely the tensor product of hermitean line bundles.
This product can be extended to include connections.

The product of hermitean line bundles can therefore be seen as induced or pulled back from the product structure on $\sfU(1)$ along the defining transition functions.%
\footnote{The set of maps from a set into a (commutative) monoid is a (commutative) monoid itself.}
Carrying this viewpoint over from the abelian group $\sfU(1)$ to the
symmetric (or abelian) 2-group $(\invs{\Hilb},\otimes)$, for two
transition line bundles $L,K \rightarrow \CU^{[2]}$ we consider the transition line bundle obtained as their product, $L \otimes K \rightarrow \CU^{[2]}$.
The tensor product of their bundle gerbe multiplications makes this product into a bundle gerbe defined over $\CU$, which we understand as the \emph{tensor product} of the two bundle gerbes.
Similarly, given two morphisms defined by functions $E,F: \CU \rightarrow \Hilb$, their tensor product yields a morphism between the tensor products of their source and target bundle gerbes.
The tensor product structure in $\Hilb$ includes a tensor product on morphisms between Hilbert spaces, and this naturally pulls back to 2-morphisms of bundle gerbes.
In this way, the symmetric monoidal structure on $(\Hilb, \otimes)$ induces
a {symmetric monoidal structure} on $\BGrb^\nabla(M)$ which we also denote by $\otimes$.%
\footnote{Functors from a category into a (symmetric) monoidal category form a (symmetric) monoidal category themselves.}

We have not yet exploited the full abelian group structure on
$\sfU(1)$; so far we have only made use of its commutative monoid structure.
The remaining structure is the existence of {inverses}.
For a line bundle defined by $g: \CU^{[2]} \rightarrow \sfU(1)$ we obtain new descent data by $g \mapsto g^{-1}$, the function $g$ composed with inversion in the group $\sfU(1)$.
The hermitean line bundle defined by this transition data is the same as the {dual} hermitean line bundle.
The line bundle $L^* \otimes L$ is {canonically isomorphic} to the trivial hermitean line bundle $I$.
For a morphism of line bundles $\psi: L \rightarrow L'$ with local
representatives satisfying $g_{ij}\, \psi_j = \psi_i\, g'_{ij}$ over
an open covering of $M$, we obtain the dual (or transpose) morphism
$\psi^\rmt: L'^* \rightarrow L^*$ by bringing the transition
functions to the opposite sides, i.e. $\psi^\rmt$ has the same local
representatives $\psi^\rmt = (\psi_i)$, but now satisfying $g_{ij}^{\prime-1}\, \psi_j = \psi_i\, g^{-1}_{ij}$.\footnote{Here we identify $\FC^* = \FC$ with pairing $\FC^* \otimes \FC \rightarrow \FC$, $w \otimes z \mapsto w \cdot z$.}

As we found in Section~\ref{sect:BGrb_Mors_loc}, an inverse (which is unique only up to unique isomorphism) of a 1-dimensional Hilbert space $V$ is given by its {dual} space $V^*$.
Thus given a bundle gerbe $\CL$ defined by a line bundle $L \rightarrow \CU^{[2]}$, consider the dual bundle $L^* \rightarrow \CU^{[2]}$.
It can be regarded as the composition of the $\invs{\Hilb}$-valued transition function $L$ with inversion in the 2-group $\invs{\Hilb}$.
The bundle gerbe multiplication on $L^*$ is then given by $\mu^{-\rmt}$, the inverse transpose of $\mu$, and the resulting bundle gerbe is called the \emph{dual bundle gerbe} $\CL^*$.
For any bundle gerbe $\CL$ there exists a {canonical} 1-isomorphism $\eta_\CL: \CI_0 \arisom \CL^* \otimes \CL$.
Starting from a morphism $(E,\alpha): \CL \rightarrow \CL'$ of bundle gerbes with isomorphisms \eqref{eq:local_mor}, we obtain its dual morphism analogously to the case of line bundles:
We just have to reinsert the original transition functions that get
cancelled out in bringing them to the opposite sides in order to then
apply $\alpha$, i.e. $(E,\alpha)^\rmt = (E,\beta): \CL'^* \rightarrow \CL^*$ is defined by
\begin{equation*}
	\xymatrixrowsep{1cm}
	\xymatrixcolsep{1.5cm}
	\xymatrix{
		L_{ij}'^* \otimes E_j \ar@{->}[d]_-{1 \otimes \mu^{-1}_{iji}} \ar@{-->}[r]^-{\beta_{ij}}& E_i \otimes L_{ij}^*\\
		L_{ij}'^* \otimes L_{ij}^* \otimes L_{ij} \otimes E_j \ar@{->}[r]_-{1 \otimes \alpha_{ij}} & L_{ij}'^* \otimes L'_{ij}\otimes E_i \otimes L^*_{ij}  \ar@{->}[u]_-{\mu_{jij}'^{-1}\otimes 1}
	}
\end{equation*}
Again, this structure is straightforwardly extended to include connections at the level of bundle gerbes as well as 1-morphisms.

\subsection{Direct sum}
\label{sect:direct_sum}

From the perspective of geometric quantisation, and also more
generally, an important feature of sections and morphisms of line bundles is that they come with an operation of a sum:
It would not be possible to construct a Hilbert space of sections otherwise.
From the point of view of local (or descent) representatives of sections of a line bundle, the additive structure is induced by the addition in the ring $\FC$.
On the category $(\Hilb, \oplus, \otimes)$, by which we are replacing $\FC$ as the ground ring, there is an additive structure $\oplus$ present.
It can readily be used to add morphisms of bundle gerbes which are
defined over a mutual covering $Y$ of $M$.
There always exists a zero 1-morphism $(E,\alpha)=0$ given by the zero vector bundle with the zero morphism.
In this case we obtain the direct sum of hermitean vector bundles over $Y$, which is compatible with connections.
The extension to general morphisms in $\BGrb^\nabla(M)$ can be found in~\cite{BSS--HGeoQuan}.

\subsection{Enrichment}
\label{sect:Structure_on_2-morphisms}

To finish off the description of the 2-category of bundle gerbes, we turn to the level of 2-morphisms.
As these are built from morphisms of twisted hermitean vector bundles with connection (at least for open coverings), they form a vector space over $\FC$.
The space of parallel sections of a vector bundle of rank~$n$ with
connection has dimension at most $n$, because every covariantly constant section spans a rank~1 sub-bundle.
Kernels of 2-morphisms have constant rank, which is a consequence of parallelity.
Because of the hermitean structure on the vector bundles, 2-morphisms also have well-defined cokernels.
With these choices of 2-morphisms, the morphism categories in $\BGrb^\nabla(M)$ are \emph{abelian categories}, just like the category $\HVBdl^\nabla(M)$.
Using the fact that on a hermitean vector bundle with connection
$\nabla$ and metric $h$ one has $\nabla h = 0$, together with parallelity of morphisms, we see that the hermitean metric evaluated on parallel morphisms is constant.
This implies that any vector space of 2-morphisms in $\BGrb^\nabla(M)$ has the structure of a finite-dimensional Hilbert space.
Composition of 2-morphisms is compatible with this
structure, as are composition and direct sum of 1-morphisms as well as
the tensor product of bundle gerbes: We say that $\BGrb^\nabla(M)$ is \emph{2-enriched in $\Hilb$}.

\section{Bundle gerbes as higher line bundles}
\label{sec:Bgerbeshigher}

\subsection{Module categories of sections}
\label{sect:Sections_of_BGrb}

From the perspective of higher geometric quantisation, and also of
higher geometry itself, we wish to find a notion of {section of a bundle gerbe}.
For a hermitean line bundle $L$ these are usually defined as maps $M \rightarrow L$ splitting the projection $\pi: L \rightarrow M$.
However, this definition is not suitable for bundle gerbes in the language employed here.
If one desires to stay in the setting of maps to a total space, the notion of space has to be generalised to incorporate total spaces of higher bundles.
A detailed treatment of this idea can be found in~\cite{NSS--principal_infty-bundles-general,NSS--principal_infty-bundles-presentations}.

There is a different, purely categorical perspective on sections of a line bundle.
Let us denote the monoidal category of hermitean line bundles on $M$ without
connection by $\HLBdl(M)$; its morphisms are given by smooth fibrewise-linear maps.
The category of hermitean line bundles with connection and parallel morphisms of line bundles is denoted $\HLBdl^\nabla(M)$.
As is evident from \eqref{eq:HLBdlmorphism}, sections of a hermitean line bundle $L \in \HLBdl(M)$ can equivalently be viewed as morphisms from the trivial hermitean line bundle $I$ to $L$,
\begin{equation*}
	\Gamma(M, L) \cong \HLBdl(M)(I,L)\ .
\end{equation*}
The trivial line bundle $I$ can in fact be {defined} to be the monoidal unit in $(\HLBdl(M),\otimes)$.

Hence for a bundle gerbe $\CL$ we define the \emph{category of sections of $\CL$} via
\begin{equation*}
	\Gamma(M,\CL) := \BGrb^\nabla(M)(\CI_0,\CL)\ .
\end{equation*}
The idea for this definition of sections of bundle gerbes is not new (cf.~\cite{Waldorf--Sections_on_mathoverflow}), but, to our knowledge, has not appeared in the literature before in this form.
Here we write the category with connections, since parallelity is
imposed at the level of 2-morphisms, whereas 1-morphisms are unconstrained; we shall require bicovariantly constant 2-morphisms in Section~\ref{sect:2Hspace_of_BGrb}.
Hence instead of the trivial bundle gerbe we must use the trivial bundle gerbe with the trivial connection for consistency.
Note that $\BGrb^\nabla(M)(\CI_0,\CI_0)$ carries both direct sum and tensor product, and that the tensor product distributes over the direct sum in a categorical sense.
Such a category is commonly referred to as a \emph{rig category}, since it has the structure of a categorified ring but is missing the additive inverses of objects.
Other important examples of rig categories include $(\HVBdl^\nabla(M), \oplus, \otimes)$ and $(\Hilb, \oplus, \otimes)$.

Every morphism category $\BGrb^\nabla(\CL,\CL'\,)$ is an abelian monoidal category under direct sum of morphisms and enriched in $\Hilb$ as discussed in Section~\ref{sec:2catBGrb}.
Under the natural isomorphisms $l_{\CL}: \CI_0 \otimes \CL \arisom \CL$ and $r_{\CL}: \CL \otimes \CI_0 \arisom \CL$, we obtain an action of the rig category $\BGrb^\nabla(M)(\CI_0,\CI_0)$ on every morphism category.
It is given (as a right-action) by
\begin{equation}\label{eq:rigaction}
	(E,\alpha) \vartriangleleft (F,\beta) = r_{\CL'} \circ \big( (E,\alpha) \otimes (F,\beta) \big) \circ r_{\CL}^{-1}\ .
\end{equation}
Hence such actions exist in particular on the categories of sections of any bundle gerbe $\CL$.
(Note that this module action extends the one obtained from composition in~\cite{Waldorf--Sections_on_mathoverflow} to morphism categories between any two bundle gerbes, rather than just being defined on sections.)
We may  use the equivalence from \eqref{eq:Reduction_functor} to translate this into an action of $\HVBdl^\nabla(M)$ on any morphism category.

The action defined by \eqref{eq:rigaction} is built from the tensor product of vector bundles and therefore distributes over the direct sum.
Hence the morphism categories have the structure of \emph{rig module categories over $\HVBdl^\nabla(M)$}.
The term `module category' usually refers to a category with an action of a monoidal category.
Here we actually have much more:
The acting category is a rig category, the categories acted upon are monoidal, and the action respects both the rig structure on the right as well as the monoidal structure on the left.
Hence this kind of action of categories is as closely related to the algebraic notion of a module over a ring as we can get.

\subsection{Hermitean 2-bundle metrics}
\label{sect:2-bdl_metric}

In this section we give a non-technical account of what we term a hermitean 2-bundle metric on a hermitean bundle gerbe.
This extends the common point of view that a bundle gerbe is a higher line bundle to the statement that a \emph{hermitean} bundle gerbe as considered here is a higher \emph{hermitean} line bundle.
A full treatment can be found in~\cite{BSS--HGeoQuan}.
The existence of a hermitean structure on the prequantum line bundle $L \to M$ is crucial in geometric quantisation, as it is this structure which allows one to naturally promote the vector space structure on sections of $L$ to a Hilbert space structure.
That is, the notion of amplitudes and probabilities, and hence the possibility of making contact with quantum mechanics in geometric quantisation relies on the existence of a hermitean metric on $L$.
Thus, it will be necessary to have a higher analogue of such a hermitean metric on the prequantum bundle gerbe $\CL$.

A hermitean line bundle is a complex line bundle with a fibrewise hermitean
metric.
On morphisms $\psi: L \rightarrow L'$ there is an induced hermitean metric.
Using the identification $C^\infty(M,\FC) \cong \HLBdl(M)(I,I)$, this turns out to be an operation in the category of line bundles on $M$,
\begin{equation*}
	h: \overline{\HLBdl(M)(L,L'\,)} \times \HLBdl(M)(L,L'\,) \longrightarrow \HLBdl(M)(I,I)\ ,
\end{equation*}
where the bar over the first argument indicates that $h$ is $C^\infty(M,\FC)$-antilinear in this argument.
To see this, we first use the hermitean structures on the objects of $\HLBdl(M)$ to obtain an antilinear isomorphism
\begin{equation*}
\begin{aligned}
	&\theta: \overline{\HLBdl(M)} \arisom \HLBdl(M)\ ,\quad \theta \big( L \overset{\psi}{\longrightarrow} L' \big) \longmapsto \big( L^* \overset{\psi^{\rmt*}}{\longrightarrow} L'^* \big)\ ,
\end{aligned}
\end{equation*}
where the superscript `t' refers to the fibrewise transpose of a morphism, and the star operation takes the fibrewise adjoint with respect to the hermitean structures.
Given a second morphism $\phi: L \rightarrow L'$ we can then use the tensor product of line bundles to obtain a morphism
\begin{equation*}
	\theta(\psi) \otimes \phi: L^* \otimes L \longrightarrow L'^* \otimes L'\ .
\end{equation*}
Recalling that $L^* \otimes L$ is {canonically} isomorphic to $I$ in $\HLBdl(M)$, we obtain an endomorphism of the trivial bundle as the composition
\begin{equation*}
	\eta_{L'}^{-1} \circ \big( \theta(\psi) \otimes \phi \big) \circ \eta_{L}  \ \in \ \HLBdl(M)(I,I)\ ,
\end{equation*}
where $\eta_L: I \arisom L^* \otimes L$ is the canonical isomorphism given by $\eta_L(z) = z\, 1_L$.
One then has
\begin{equation}
\label{eq:bdl_metric_from_tensor_produc}
	\eta_{L'}^{-1} \circ \big( \theta(\psi) \otimes \phi \big) \circ \eta_{L} = h(\psi, \phi)
\end{equation}
under the identification $\HLBdl(M)(I,I) \cong C^\infty(M,\FC)$.
Thus the hermitean metrics on line bundles can equivalently be obtained via the Riesz-type equivalence $\theta$ which acts $C^\infty(M,\FC)$-antilinearly on morphisms.

This functorial version of a hermitean bundle metric is easier to lift to the higher categorical world of bundle gerbes.
We have already found a dual (or transpose) of a morphism between bundle gerbes in Section~\ref{sect:pullproddual}, but the ordinary dual functor produces morphisms going in the wrong direction and is linear rather than antilinear.
However, given a bundle gerbe morphism $(E,\alpha): \CL \rightarrow \CL'$, observe that
\begin{equation*}
	(E^*, \alpha^{-\rmt}) = (E^*, \alpha^{\rmt*}): \CL^* \longrightarrow \CL'^*\ ,
\end{equation*}
just as desired.
From the point of view of local representations of a section this amounts to replacing the local representatives by their complex conjugates.
In particular, if we just consider a hermitean vector bundle $E\to M$ as an endomorphism $(E,1_E): \CI_0 \rightarrow \CI_0$ of the trivial bundle gerbe on $M$, this construction yields the dual vector bundle.\footnote{Because of the hermitean structure, the dual $E^*$ is canonically isomorphic to the complex conjugate of $E$.}
Thus we can define a 2-functor
\begin{equation*}
	\Theta: \BGrb^\nabla(M) \longrightarrow \BGrb^\nabla(M) 
\end{equation*}
acting on 1-morphisms as
\begin{equation*}	
	\Theta\big( (E,\alpha): \CL \rightarrow \CL' \big) = \big( (E^*,\alpha^{\rmt*}): \CL^* \rightarrow \CL'^* \big)\ .
\end{equation*}
On 2-morphisms we could just have it take the transpose, thereby acting contravariantly on 2-morphisms.
However, as we wish to employ the hermitean structure we define the action on $\psi: (E,\alpha) \Rightarrow (F,\beta)$ by
\begin{equation*}
	\Theta (\psi) = \psi^{\rmt*} = \psi^{*\rmt}: \Theta(E,\alpha) \Longrightarrow \Theta(F,\beta)\ .
\end{equation*}
With this convention, $\Theta$ becomes a covariant 2-functor,\footnote{It is also compatible with pullbacks, whence it is a transformation of 2-stacks.}
which is compatible with tensor products and direct sums of morphisms, and it acts antilinearly in the sense that $\Theta((E,\alpha) \vartriangleleft F) = \Theta(E,\alpha) \vartriangleleft F^*$ for any hermitean vector bundle $F \in \HVBdl^\nabla(M)$ and morphism of gerbes $(E,\alpha)$.

Given a pair of morphisms $(E,\alpha), (F,\beta): \CL \rightarrow \CL'$ we can now mimic~\eqref{eq:bdl_metric_from_tensor_produc}:
We define
\begin{equation*}
	\frh \big( (E,\alpha), (F,\beta) \big) := \rmR \big( \eta_{\CL'}^{-1} \circ \big( \Theta(E,\alpha) \otimes (F,\beta) \big) \circ \eta_{\CL} \big) \ \in \  \HVBdl^\nabla(M)\ .
\end{equation*}
We included the equivalence $\rmR$ from \eqref{eq:Reduction_functor} in order to end up in the more familiar category of hermitean vector bundles on $M$, which corresponds directly to functions on $M$ valued in $\Hilb$; they can be regarded as \emph{higher functions}.
Without $\rmR$ in the definition of $\frh$ we would only map to the endomorphisms of the trivial bundle gerbe.\footnote{These endomorphisms can analogously serve as higher functions, but $\HVBdl^\nabla(M)$ is equivalent as a rig category and more convenient to work with.}
Via the above module actions, morphisms of bundle gerbes, and in particular sections of bundle gerbes, carry a module structure over these higher functions, just as morphisms and sections of vector bundles are modules over the ring of functions $C^\infty(M,\FC)$.
The functor $\frh$ is sesquilinear with respect to this module action and complex conjugation of hermitean vector bundles as defined by $\Theta$.
Thus we obtain a \emph{2-bundle metric} on $\BGrb^\nabla(M)(\CL, \CL'\,)$ as the composition of functors and natural 1-isomorphisms inherent to the category of bundle gerbes with connection on $M$.
It has the useful property%
	\footnote{This is true with our convention of taking 2-morphisms to be {parallel}.
	Without this feature the isomorphism \eqref{eq:GammaparBGrb} still holds if one omits the subscript `par'.}
\begin{equation}\label{eq:GammaparBGrb}
	\Gamma_{\rmpar} \big( M,\, \frh \big( (E,\alpha), (F,\beta) \big) \big) \cong \BGrb^\nabla(M) \big( (E,\alpha), (F,\beta) \big)\ ,
\end{equation}
where $\Gamma_{\rmpar}(M,-): \HVBdl^\nabla(M) \rightarrow \Hilb$ is
the functor that assigns to a hermitean vector bundle on $M$ with
connection its Hilbert space of covariantly constant sections; this follows from the fact that $\Theta(E,\alpha) \otimes (F,\beta)$ is descent data for a hermitean vector bundle with connection on $M$, while the 2-morphisms between these bundles precisely correspond to sections of this descent bundle.
With our choice of $\Theta$, however, the isomorphism \eqref{eq:GammaparBGrb} holds on objects only.
A natural isomorphism exists only in the form
\begin{equation*}
	\eta_{(E,\alpha),(F,\beta)}:\ \BGrb^\nabla(M) \big( (E,\alpha), (F,\beta) \big)\ \Arisom\ \frh \circ \big( \vartheta(E,\alpha) \times (F,\beta) \big)\ ,
\end{equation*}
where $\vartheta: \BGrb^\nabla(M) \rightarrow \BGrb^\nabla(M)$ is the 2-functor which is the identity on objects and 1-morphisms, but sends 2-morphisms to their adjoints, i.e. $\vartheta(\psi) = \psi^*$.

\section{2-Hilbert spaces}
\label{sect:2-HSpaces}

The key idea of the preceding sections has been that the ground ring $(\FC,+,\cdot\,)$ with its natural structure of a finite-dimensional Hilbert space should be replaced by a categorical object.
This object should be endowed with categorical analogues of the algebraic structures of $\FC$, including the inner product.
General accounts can be found for example in~\cite{Baez--2-Hilbert_spaces,KV--2-Cats_and_Zam_eqns}.

There are (at least) two candidate categories which provide categorifications of $\FC$.
The complex numbers form a set, or a 0-category.
From any set $S$ one can canonically construct a category by adding to the data an identity morphism for each element of $S$.
This is called the \emph{discrete category $\disc(S)$ of the set $S$}.\footnote{Recall that we have already encountered a higher categorical version of this construction in Section~\ref{sect:Rmks_on_global_picture}.}
In our case, $\disc(\FC) = (\FC \rightrightarrows \FC)$; source and target of $z \in \FC$ are both $z$.
Composition in this category is trivial since all morphisms are identities.
The set of morphisms inherits the ring structure from the objects.
From this construction we therefore obtain a category in $\Hilb$.
This means that both collections of objects and morphisms form finite-dimensional Hilbert spaces, and the structure maps are compatible with this structure.

The category $\disc(\FC)$ carries a natural action of the discrete category $\disc(\sfU(1)) = (\sfU(1) \rightrightarrows \sfU(1))$.
However, this is not the categorification of the group $\sfU(1)$ which appears in the definition of a bundle gerbe, where the pertinent 2-group is $(\invs{\Hilb},\otimes)$.
For every 1-dimensional Hilbert space there exists a unitary isomorphism to $\FC$, and endomorphisms of any such Hilbert space identify canonically with $\FC$.
Thus the inclusion
\begin{equation}\label{eq:BU1Hilb}
\xymatrix{
	\scB\sfU(1) = \big(\sfU(1) \rightrightarrows * \big) \ \ar@{^{(}->}[r] \ & \  (\invs{\Hilb}, \otimes) \ ,
	}
\end{equation}
sending the single object of the suspension $\scB\sfU(1)$ of $\sfU(1)$ to $\FC$, is an equivalence of categories.
It is compatible with the tensor structure and therefore is an equivalence of 2-groups.
This relates our definition of bundle gerbes to the common statement that they are principal 2-bundles with structure 2-group $\scB\sfU(1)$.
From a topological perspective, this structure 2-group is big enough to capture all bundle gerbes up to stable isomorphism.\footnote{In our language, an isomorphism is \emph{stable} if it uses the coarsest common refinement of the coverings that the bundle gerbes are defined over.}
However, it does not capture important examples such as the central extensions of loop groups discussed in Section~\ref{sect:Bundle_gerbes}.
For this reason we stick to our choice of $(\invs{\Hilb},\otimes)$ as the structure 2-group.\footnote{One might also call $(\invs{\Hilb},\otimes)$ a different model for $\scB\sfU(1)$.}

The most natural object to represent $(\invs{\Hilb},\otimes)$ on is the rig category $(\Hilb, \oplus, \otimes)$.
It acts on itself via the tensor product in the same way that a ring acts on itself via multiplication.
In this way, $\Hilb$ is a module category over itself generated by a single object.
As in the case of the inclusion \eqref{eq:BU1Hilb}, there exist unitary isomorphisms for every Hilbert space of dimension $n$ to $\FC^n$ with its natural inner product.
A choice of two such isomorphisms establishes a bijection between morphisms $V \rightarrow W$ and spaces of complex $n{\times}m$-matrices $\Mat( n{\times}m, \FC)$.
Hence the inclusion
\begin{equation}\label{eq:MatHilb}
\xymatrix{
	\big( \Mat( -{\times}-, \FC) \rightrightarrows \NN_0 \big) \ \ar@{^{(}->}[r] \ & \ \Hilb
	}
\end{equation}
is an equivalence of rig categories, where the category on the left has objects given by numbers $n \in \NN_0$ and morphisms $n \rightarrow m $ given by $n{\times}m$-matrices.
Using the smaller category in the associated construction of morphisms from Section~\ref{sec:2catBGrb} would only enable us to obtain trivial hermitean vector bundles $E \rightarrow Y$, and thus only a restricted class of morphisms of bundle gerbes.
If we would like to understand a hermitean vector bundle on $Y$ as a map into a generalised ring, then this generalised ring has to be the bigger category $\Hilb$ because although the fibres of any rank~$n$ hermitean vector bundle are all isomorphic to $\FC^n$, they are {not $\FC^n$ on the nose}:
we can find a consistent {simultaneous} identification of all fibres with $\FC^n$ if and only if the vector bundle is trivialisable.
This justifies our use of $(\Hilb, \oplus, \otimes)$ as the ground rig.

The space of smooth sections of a line bundle gives rise to a pre-Hilbert space, which is algebraically a $\FC$-module together with a sesquilinear $\FC$-valued inner product.
If the underlying vector space of such a module is of finite dimension, then a pre-Hilbert space is automatically a Hilbert space.
Translating this to the setting where the ground rig is $\Hilb$, we are interested in {rig-module categories over $\Hilb$}.
Such a category is a symmetric monoidal category $(\scC, \oplus)$ together with a module action of $\Hilb$,\footnote{We could probably drop the symmetric condition, at least at this stage, but we keep it here for convenience.} that we may choose to be from the right, which is a functor $\vartriangleleft: \scC \times \Hilb \rightarrow \scC$ satisfying the categorified properties of a ring module action up to coherent isomorphisms.
Additionally, we would like the action of the functor on morphisms to be compatible with the vector space structures on the morphism spaces in $\Hilb$.
The action $C \mapsto C \vartriangleleft \FC \cong C$ of the unit element in $\Hilb$ then forces all the morphism sets in $\scC$ to be vector spaces, i.e. we demand that $\scC$ be enriched in $\Vect$, the category of finite-dimensional complex vector spaces.\footnote{Recall from Section~\ref{sect:Structure_on_2-morphisms} that this means each set of morphisms $\scC(a,b)$ is a finite-dimensional vector space, and composition is compatible with this structure.}
Sometimes $\scC$ is required to be enriched in $\Hilb$, but it seems more natural to just require enrichment in $\Vect$ for the general case.\footnote{If we were to extend this theory to encompass also infinite-dimensional Hilbert spaces, then the ground rig $\Hilb_\infty$ of possibly infinite-dimensional Hilbert spaces with continuous operators would be enriched in $\Vect_\infty$, the possibly infinite-dimensional vector spaces, rather than in $\Hilb_\infty$. There is still however additional structure on the morphism spaces: they are Banach spaces.}
A \emph{2-vector space} is a monoidal $\Vect$-enriched category $\scC$ with a $\Hilb$ rig-module action that factors through the forgetful functor $\Hilb \rightarrow \Vect$ which drops the inner product.

We are interested in \emph{2-Hilbert spaces}.
In order to obtain a Hilbert-like structure we need an inner product.
Since our ground rig is $\Hilb$, this should be the target of inner products.
Thus we define an inner product on a 2-vector space $\scC$ to be a sesquilinear, non-degenerate functor $\< -,- \>_\scC : \scC \times \scC \rightarrow \Hilb$.
Sesquilinearity is to be understood with respect to the rig-module action of $\Hilb$ and the involution on $\Hilb$ given by
\begin{equation*}
	\Theta: \Hilb \longrightarrow \Hilb \ ,\, \quad
	\big( V \overset{\psi}{\longrightarrow} V' \, \big)\longmapsto \big( V^* \overset{\psi^{\rmt*}}{\longrightarrow} V'^* \big)\ ,
\end{equation*}
where as in Section~\ref{sect:2-bdl_metric} we use $\Theta(\psi) = \psi^{\rmt*}$.
Non-degeneracy here means that $\< V,V \>_{\scC} = 0$ if and only if $ V = 0$ in $\scC$.\footnote{In a rig (and even generally in a ring) there is no notion of positive definiteness.}
The category $\Hilb$ itself carries such an inner product given by
\begin{equation*}
	\< V, W \>_{\Hilb} = \Theta V \otimes W = V^* \otimes W \cong \Hilb(V,W)\ ,
\end{equation*}
where the last identification follows since we are considering finite-dimensional Hilbert spaces.
Given two morphisms $\psi: V \rightarrow V'$ and $\phi: W \rightarrow W'$, we obtain
\begin{equation*}
	\< \psi, \phi \>_{\Hilb} = \Theta(\psi) \otimes \phi: \<V,
        W\>_{\Hilb} \longrightarrow \<V', W'\, \>_{\Hilb}\ .
\end{equation*}
 This is the same as the 2-bundle metric from Section~\ref{sect:2-bdl_metric} on the category $\HVBdl^\nabla(\pt) = \Hilb$.
 With the definition $\<-,-\>_{\Hilb} = \Theta (-) \otimes (-)$, it is not true that $\<-,-\>_{\Hilb}$ is the same as $\Hilb(-,-)$ for the same reasons as explained at the end of Section~\ref{sect:2-bdl_metric}:
 This identity holds on objects only.
 The inner product functor is covariant in both arguments, whereas the hom-functor is contravariant in its first argument.
In particular, for $\psi, \phi \in \Hilb(\FC,\FC)$ we have $\< \psi, \phi \>_{\Hilb} = \< \psi, \phi \>_\FC$ under the identification $\Hilb(\FC,\FC) \cong \FC$.
Thus the original Hilbert space $\FC$ (also as an algebra with inner product) sits inside $\Hilb$ as the endomorphism space of the monoidal unit.
Using $\Hilb(-,-)$ as the inner product would only yield $\Hilb(\psi,\phi) = \psi\, \phi$.

Morphisms of 2-Hilbert spaces are functors of $\Hilb$-module categories, i.e. functors compatible with the module actions and the direct sums.
As morphisms are certain functors, there exists a notion of 2-morphism of 2-Hilbert spaces given by natural transformations of $\Hilb$-module functors.
By including morphisms and 2-morphisms of 2-Hilbert spaces, we obtain a {2-category $2\Hilb$ of 2-Hilbert spaces}.

A morphism of 2-Hilbert spaces is \emph{unitary} if it preserves the inner product up to a well-defined natural isomorphism.
The simplest type of 2-Hilbert spaces are free 2-Hilbert spaces, i.e. those of the form $\Hilb^n$.
Given a generic 2-Hilbert space $\scC$, there might be unitary isomorphisms $U: \scC \arisom \Hilb^n$ for some $n \in \NN_0$.
On objects only, we then have
\begin{equation*}
	\scC(C,D) \cong \Hilb^n(UC, UD) = \bigoplus_{i,j=1}^n\,
        \Hilb\big((UC)_i, (UD)_j\big)\ \in\ \Mat(n{\times}n,\Hilb)\ . 
\end{equation*}
Because of the unitarity of $U$, we therefore have
\begin{equation*}
	\< C, D \>_{\scC} \cong \< UC, UD \>_{\Hilb^n} = \bigoplus_{i=1}^n\, \big\< (UC)_i, (UD)_i \big\>_{\Hilb} = \tr \big( \Hilb^n(UC, UD) \big) \ \in \ \Hilb\ .
\end{equation*}
This is completely analogous to what happens in an ordinary Hilbert space $V$:
Given two vectors $\psi, \phi \in V$, we can view them as morphisms $\psi, \phi \in \Hilb(\FC,V)$, and combine them into a morphism $\phi \circ \psi^* \in \Hilb(V,V)$.
With this identification of vectors, given any unitary transformation $u: V \rightarrow \FC^n$ we obtain an endomorphism of $\FC^n$ as
\begin{equation*}
	u \circ (\phi \circ \psi^*) \circ u^{-1} = u(\phi) \circ u(\psi)^* \ \in \  \Hilb(\FC^n,\FC^n) = \Mat(n{\times}n,\FC)\ ,
\end{equation*}
where $u(\phi)= u \circ \phi $, and finally
\begin{equation*}
	\< \psi, \phi \>_V = \tr \big( u(\phi) \circ u(\psi)^* \big) = \tr( \phi \circ \psi^*)\ .
\end{equation*}
The transformation and invariance laws for these objects translate to 2-Hilbert spaces with equalities weakened to canonical isomorphisms.

To summarise, we take a 2-Hilbert space to be a $\Vect$-enriched rig-module category $\scC$ over $\Hilb$ with a covariant inner product functor $\< -,- \>_{\scC}: \scC \times \scC \rightarrow \Hilb$.
An object $C$ in a 2-Hilbert space $\scC$ is called \emph{simple} if $\scC(C,C) \cong \FC\, 1_C$, i.e. its endomorphism algebra is 1-dimensional, and \emph{normalised} or \emph{of unit length} if $\< C,C \>_{\scC} \cong \FC$.
If we are in the situation that $\<-,-\>_{\scC}$ and $\scC(-,-)$ agree on objects, the notions of simplicity and unit length coincide.
Particularly nice 2-Hilbert spaces are those which are \emph{semisimple}, i.e. where every object decomposes into direct sums of simple objects (up to isomorphism), and where we additionally assume that the inner product agrees with the hom-functor on objects.
In such a decomposition, several isomorphic copies of the same simple object may occur.
Fixing one simple object $C_i$ in each isomorphism class of simple objects, each of the $C_i$ will occur $n_i$ times for some $n_i \in \NN_0$.
Hence we can write $C = \bigoplus_i\,  (C_i \vartriangleleft \FC^{n_i})$.
Therefore any semisimple 2-Hilbert space is unitarily equivalent to a free 2-Hilbert space via the choice of an orthonormal basis, which in this case amounts to picking one object in each isomorphism class of simple objects.

There does not yet exist a generally accepted notion of a 2-Hilbert space.
The definition we have presented here may be altered in several ways.
To point out just a few variations, sometimes the inner product is assumed to be given by the hom-functor, sometimes the underlying category is assumed to be abelian or semisimple, or $\Hilb$-enrichment may be required instead of $\Vect$-enrichment.
An extensive treatment can be found in~\cite{Baez--2-Hilbert_spaces}.
Another interesting question is of the necessity of a 2-Hilbert space completion.
This is completely open as of yet; some approaches to parameterised families of Hilbert spaces can be found in~\cite{CS--2-categorical_reps_and_state_sum_application,CY--Measurable_Cats_and_2groups}, and a self-contained account in view of general continuous 2-group representations is given in~\cite{BBFW--Infinite-dimensional_reps_of_2-groups}.

The prototypical 2-Hilbert space, which is  semisimple, is $\Hilb$ itself, just as $\FC$ is the prototypical Hilbert space.
The question is which of the features of $\FC$ to take over to the general setting since $\FC$ is an extremely special object with a particularly rich structure.
Presumably, the first step of categorification from objects based on sets to objects based on categories is the hardest to find, whereas one might hope that once the first additional layer and weakening of structure has been found it is easier to see how even higher $n$-Hilbert spaces should be defined.
In Section~\ref{sect:2Hspace_of_BGrb} we will encounter an example of a 2-Hilbert space, in the above sense, arising from the geometric data of a bundle gerbe.

\section{The prequantum 2-Hilbert space of a bundle gerbe}
\label{sect:2Hspace_of_BGrb}

In this section we will put to use the properties and structures of bundle gerbes that we have accumulated in earlier sections to the goal of obtaining a 2-Hilbert space of sections of a bundle gerbe on a manifold $M$.
This extends the statement in~\cite{Waldorf--Sections_on_mathoverflow} that $\Gamma(M,\CL)$ is a 2-vector space.
If the category of sections of $\CL$ is to be the space of states of a quantum theory, there has to be a pairing on these states in order to define amplitudes (whose interpretation in physical terms is yet to be understood in the higher settings) and hermitean observables.
Using sections of a bundle gerbe as higher quantum states has been proposed in~\cite{Rogers--Thesis,Rogers--HGeoQuan_talk} and also in~\cite{Schreiber--2-states_talk}, though the hermitean and inner product structures have not been present there, and the treatment was restricted to local bundle gerbes.

Recall from Section~\ref{sect:GeoQuan_review} that in geometric quantisation the pre-Hilbert space of interest was the space of sections $\Gamma(M,L)$ with the inner product given by integrating the evaluation of the hermitean metric of $L$ on a pair of sections over $M$.
In 2-plectic quantisation we would now like to replace the hermitean line bundle $L$ by a hermitean bundle gerbe $\CL$.
In Section~\ref{sect:Sections_of_BGrb} we defined the category of sections of $\CL$ as $\Gamma(M,\CL) = \BGrb^\nabla(M)(\CI_0,\CL)$.
We also elaborated that this forms a module category over the rig category $\HVBdl^\nabla(M)$ of higher functions on $M$.
From here on, we assume $M$ to be connected.

For the Hilbert space structure of $\Gamma(M,L)$ the relevant module structure is that over $\FC$ rather than that over $C^\infty(M,\FC)$.
However, the former module structure is actually contained in the latter as follows.
We simply note that the action of a complex number $z \in \FC$ on a section $\psi \in \Gamma(M,L)$ factors through the inclusion of complex numbers as constant functions,
\begin{equation*}
\xymatrix{
	c: \FC \ \ar@{^{(}->}[r] \ & \ C^\infty(M,\FC) \ , \quad
	c(z) (x) = z \ ,
	}
\end{equation*}
for all $x\in M$.
For a 2-Hilbert space we need a module structure over $\Hilb$.
The above picture of the ground ring sitting inside the ring of functions as constant functions generalises appropriately:
Given a Hilbert space $V \in \Hilb$ we can form the trivial hermitean vector bundle with fibre $V$, i.e. $M {\times} V \in \HVBdl(M)$.
This becomes even clearer from the perspective that hermitean vector bundles are functions into $\Hilb$: the bundle $M {\times} V$ is precisely the constant function $x \mapsto V$.
However, this is not yet a higher function as was defined in Section~\ref{sect:2-bdl_metric} since we are missing a connection.
Again there is a constant choice for a connection on a trivial bundle, namely the trivial connection given by just the exterior derivative.
Thus the inclusion of higher numbers into higher functions reads as
\begin{equation*}
\xymatrix{
	c: \Hilb \ \ar@{^{(}->}[r] \ & \  \HVBdl^\nabla(M) \ , \quad
	V \longmapsto \big( M {\times} V,\, \dd\, \big)\ .
	}
\end{equation*}
This is  a functor, and it is compatible with the structure functors $\oplus$, $\otimes$ and $\Theta$ on $\Hilb$.
Composing the module action of $\HVBdl^\nabla(M)$ with this inclusion yields a module action of $\Hilb$ on $\Gamma(M,\CL)$ for any $\CL \in \BGrb^\nabla(M)$.

For the inner product, we have already constructed a 2-bundle metric on $\CL$ in the algebraic sense in Section~\ref{sect:2-bdl_metric}.
This was defined on sections as a functor $\frh: \Gamma(M,\CL) \times \Gamma(M,\CL) \rightarrow \HVBdl^\nabla(M)$.
In the case of a hermitean line bundle, the inner product of two sections is given by first inserting them into the metric and then integrating the resulting function over $M$.
Inserting two sections into the 2-bundle metric yields a hermitean vector bundle with connection.
We would now like to integrate this over $M$, but first we have to settle on a suitable notion of integration.

First of all, viewing the bundle as a function to $\Hilb$ would suggest adding up all the fibres, possibly weighted with a measure.
The measure should then be valued in $\Hilb$ as well, but such a
measure would have that be of rank~1 at least if it is non-vanishing,
leading inevitably to infinite-dimensional Hilbert spaces which are
not separable.
Another approach is known from harmonic analysis (see for instance~\cite{Folland--Harmonic_analysis}), where the \emph{direct integral} of a family of Hilbert spaces over $M$ (e.g. a hermitean vector bundle with possibly infinite-dimensional fibres) is defined to be the space of $\rmL^2$-sections of the family.
We could technically apply the direct integral to our finite-rank
hermitean vector bundles, but this would again produce
infinite-dimensional Hilbert spaces, which are now separable.
From the point view of our ground rig $\Hilb$, an infinite-dimensional separable Hilbert space $\CH$ would be the same as $\infty$ is for $\FC$; for example, $V \oplus \CH \cong \CH$ and $V \otimes \CH \cong \CH$ for any Hilbert space $V$ (including $V = \CH$).

A modification of the direct integral which allows us to stay in the finite-dimensional world is the functor
\begin{equation*}
	\smint^{\rmpar}_M = \Gamma_{\rmpar}(M,-):\ \HVBdl^\nabla(M) \longrightarrow \Hilb \ , \quad
	E \longmapsto \smint^\rmpar_M\, E = \Gamma_{\rmpar}(M,E)\ .
\end{equation*}
This respects the direct sum and involution in $\HVBdl^\nabla(M)$, and as a consequence is compatible with the module action of $\Hilb$, i.e.
\begin{equation*}
\begin{aligned}
	\smint^{\rmpar}_M\, \big( (E \vartriangleleft V) \oplus (F^* \vartriangleleft W) \big) &= \Gamma_{\rmpar} \big( M, (E \vartriangleleft V) \oplus (F^* \vartriangleleft W) \big)\\[4pt]
	&= \big( \smint^{\rmpar}_M\, E \big) \otimes V\ \oplus\ \big( \smint^{\rmpar}_M\, F \big)^* \otimes W\ .
\end{aligned}
\end{equation*}
The inner product on $\Gamma_{\rmpar}(M,E) \in \Hilb$ is canonically obtained from the hermitean metric $h$ on $E$:
For $\psi, \phi \in \Gamma_{\rmpar}(M,E)$ define their inner product to be $h_{|x}(\psi, \phi) \in \FC$ for {any} $x \in M$.
The independence of the choice of point $x$ follows from the fact that
the sections and the hermitean metric are (bi)covariantly constant, and the assumption that $M$ be connected.

Therefore to a bundle gerbe $\CL \in \BGrb^\nabla(M)$ with connection on $M$, we obtain a 2-Hilbert space of sections given as
\begin{equation*}
	\scH(\CL) = \big( \Gamma(M,\CL),\, \smint^{\rmpar}_M \circ\, \frh \big) \ \in \ 2\Hilb\ .
\end{equation*}
It follows immediately from the functorial nature of this assignment that we have in fact found a 2-functor
\begin{equation*}
	\scH: \BGrb^\nabla(M) \longrightarrow 2\Hilb\ .
\end{equation*}
It is merely the global section functor $\Gamma(M,-) = \BGrb^\nabla(M)(\CI_0,-)$ with the additional information of the inner product; this matches the situation for line bundles precisely.
This inner product is non-degenerate since for every vector bundle with connection, the identity morphism is parallel:
Hence $\big\< (E,\alpha), (E,\alpha) \big\>_{\Gamma(M,\CL)} = 0$ if and only if $(E,\alpha) = 0$.
Thus, $\scH$ refines the global section functor $\Gamma(M,-)$ to a functor from bundle gerbes to 2-Hilbert spaces.
This is in analogy to how sections of a hermitean line bundle define not just a vector space, but even a Hilbert space of sections, and it is this \emph{Hilbert} space structure that makes the space of sections of a prequantum line bundle into a viable habitat for a quantum theory as it allows to define transition amplitudes and self-adjoint observables.

Let us point out a few more features of the parallel section functor.
If $E \in \HVBdl^\nabla(M)$, the assignment $U \mapsto \Gamma_{\rmpar}(U,E_{|U})$ produces a sheaf of Hilbert spaces on $M$.
We furthermore get a functor
\begin{equation*}
	\Gamma_{\rmpar}: \HVBdl^\nabla(M) \longrightarrow \Sh_{\Hilb}(M) 
\end{equation*}
from the category of hermitean vector bundles with connection on $M$
to the category $\Sh_{\Hilb}(M) $ of $\Hilb$-valued sheaves on $M$.
Every $\Gamma_{\rmpar}(-,E)$ is a sheaf of modules over the sheaf of constant functions $\FC_M \cong \Gamma_{\rmpar}(-, c(\FC))$.
There is a generalised Mayer-Vietoris property for open sets $U,U' \subset M$, as there is an exact sequence
\begin{equation}
\label{eq:parint_exact_seqence}
	\xymatrix{
		0 \ar@{->}[r] & \smint^{\rmpar}_{U \cup U'}\, E \ar@{->}[r] & \smint^{\rmpar}_U\, E\ \oplus\ \smint^{\rmpar}_{U'}\, E \ar@{->}[r] & \smint^{\rmpar}_{U \cap U'}\, E
	}\ .
\end{equation}
This sequence does not extend to a short exact sequence for general $U$ and $U'$.
For example, consider $U$ and $U'$ with non-trivial fundamental group, but such that their intersection is contractible.
Let $E$ be a flat hermitean line bundle with non-trivial holonomy around the cycles in $U$ and $U'$.
Then the first and second Hilbert spaces of the exact sequence~\eqref{eq:parint_exact_seqence} are zero, while the rightmost space is isomorphic to $\FC$.
Nevertheless, this exact sequence resembles the familiar property of
the Lebesgue integral: $\int_{U \cup U'}\, f\, \dd\mu = \int_U\, f\, \dd\mu\ +\ \int_{U'}\, f\, \dd\mu\ -\ \int_{U \cap U'}\, f\, \dd\mu$.

Recall that there are two ways to obtain a Hilbert space from a pair $(E,\alpha)$ and $(F,\beta)$ of sections of $\CL$.
We can either use the inner product on the 2-Hilbert space $\scH(\CL)$ of sections of $\CL$, giving $\Gamma_{\rmpar}(M, \frh((E,\alpha),(F,\beta)))$, or we can consider the space of 2-morphisms $\BGrb^\nabla(M)((E,\alpha),(F,\beta))$ between the two sections in the 2-category of bundle gerbes.
As pointed out in Section~\ref{sect:2-bdl_metric} these two Hilbert spaces are canonically isomorphic by construction.
The corresponding isomorphisms assemble into a natural isomorphism
\begin{equation*}
	\eta'_{(E,\alpha),(F,\beta)}:\ \BGrb^\nabla(M) \big( (E,\alpha), (F,\beta) \big)\ \Arisom\ \big\< \vartheta(E,\alpha) ,\, (F,\beta) \big\>_{\Gamma(M,\CL)}\ .
\end{equation*}
From the fact that this natural isomorphism is the identity on
objects and since the morphism categories are abelian, we infer that $\Gamma(M,\CL)$ is a semisimple 2-Hilbert
space.

We summarise the most important structures and analogies in higher prequantisation with the following table (see~\cite{BSS--HGeoQuan} and Section~\ref{sec:open} for further discussion):

\begin{center}
\begin{tabularx}{\textwidth}{| p{4.5cm} | p{4.5cm} | X |}
	\hline
	{\sl Structure}	& Geometric quantisation & 2-plectic quantisation \\ \hhline{|=|=|=|}
	{\sl Geometric data} & Symplectic form $\omega$ & 2-plectic form $\varpi$ \\ \hline
	{\sl Chern-Weil object} & Line bundle $L$ & Bundle gerbe $\CL$ \\ \hline
	{\sl Ground ri(n)g} & $(\FC, +, \cdot\,)$ & $(\Hilb, \oplus, \otimes)$ \\ \hline
	{\sl Functions on $M$} & $C^\infty(M,\FC)$ & $\HVBdl^\nabla(M)$ \\ \hline
	{\sl Algebra} & Poisson algebra & unknown for full category \\ \hline
	{\sl Sections $\Gamma(M,-) =$} & $\HLBdl(M)(I,L)$ & $\BGrb^\nabla(M)(\CI_0,\CL)$ \\ \hline
	{\sl Sections form a:} & $\FC$-Hilbert space $\CH(L)$ & $\Hilb$-module category $\scH(\CL)$ \\ \hline
	{\sl Inner product} & $\int_M \, \frac{\omega^n}{n!} \circ h : \CH(L)^2 \rightarrow \FC$ & $\int^{\rmpar}_M\, \circ\, \frh: \scH(\CL)^2 \rightarrow \Hilb$ \\ \hline
	{\sl Prequantisation map} & Kostant-Souriau map & unknown \\ \hline
	{\sl Polarisation} & Lagrangian foliation & unknown \\ \hline
\end{tabularx}
\end{center}

\section{Examples}
\label{sec:Examples}

\subsection{$H$-flux on $\FR^3$}
\label{sect:R3}

In Type~II string theory, the prototypical example of a non-geometric flux background arises in the T-duality chain that starts from a 3-torus with constant geometric $H$-flux. In many situations local considerations suffice, so that one can take the decompactification limit and work on $\FR^3$ instead, see e.g.~\cite{BDL--NonAsso,Mylonas:2012pg,Aschieri-Szabo--NonAsso}. This setting is appropriate for our needs as well: the constructions of this paper apply to bundle gerbes on $\FR^3$ and we can explicitly construct the pertinent 2-Hilbert space in 2-plectic quantisation.

On $M=\FR^3$ we can consider the trivial bundle gerbe $\CL = \CI_\rho$ with non-trivial curving given by $\rho = \frac{\pi}{3}\, \epsilon_{ijk}\, x^i\, \dd x^j \wedge \dd x^k$.
It has $H$-flux $H = \dd \rho = 2\pi\, \dd x^1 \wedge \dd x^2 \wedge \dd x^3$.
Technically any curving would be admissible at this stage, but the
3-form $H$-flux we obtain from this choice shows that $\CI_\rho$ is a
geometric representative of the canonical 2-plectic form $\varpi = \dd
x^1 \wedge \dd x^2 \wedge \dd x^3$ on $\FR^3$; the corresponding
classical 3-brackets define the standard Nambu-Poisson structure on
$\FR^3$ and hence our construction will produce the prequantum
2-Hilbert space for the elusive quantised Nambu-Heisenberg algebra.
As the bundle gerbe is trivial, its category of sections is
\begin{equation*}
	\Gamma(\FR^3,\CI_\rho) \cong \BGrb^\nabla(\FR^3)(\CI_0,\CI_0) \cong \HVBdl^\nabla(\FR^3)\ .
\end{equation*}
The first step makes use of the fact that morphisms of bundle gerbes do not see their curvings,%
\footnote{This is true at least for the definition of morphisms of bundle gerbes employed here.
In other places, as e.g.~\cite{Waldorf--More_morphisms}, an additional condition is used, relating the trace of the curvature of the 1-morphism to the curvings of the source and target gerbes.
However, mathematically, as well as in view of the DBI-action of D-branes, this condition appears to be unnatural.}
while the second step is the by now familiar equivalence of categories coming from the fact that vector bundles with connection form a stack.

Here we are in the special situation that $\FR^3$ is contractible so that every hermitean vector bundle is trivialisable.
This is true, however, only at the level without connections.
Thus every hermitean vector bundle of rank~$n$ with connection is
isomorphic, as a vector bundle with connection, to $(\FR^3 {\times}
\FC^n,\, \dd + \iu\, A\, )$ for some 1-form $A \in \Omega^1(\FR^3,
\fru(n))$; we can use just $A$ to label a section of this form.
A 2-morphism $\psi: A \Rightarrow A'$ is a function $f: \FR^3
\rightarrow \Mat(n{\times}n', \FC)$ which is bicovariantly constant:
$\dd f + \iu\,A'\, f - \iu\, f\, A = 0$.
Denoting by $\Omega^1(\FR^3, \fru)$ the category with objects given by
1-forms on $\FR^3$ valued in $\fru(n)$ for some $n \in \NN_0$ and
morphisms given by bicovariantly constant functions on $\FR^3$ valued in $\Mat(n {\times} n', \FC)$ for $n,n' \in \NN_0$, we have a further equivalence of categories
\begin{equation*}
	\Gamma(\FR^3,\CI_\rho) \cong \Omega^1(\FR^3, \fru)\ .
\end{equation*}
Direct sum and tensor product on $\Omega^1(\FR^3, \fru)$ are induced
by those of connection 1-forms, so that $\oplus$ sends $(A,A'\,)$ to
$A\oplus A'$ while $\otimes$ sends $(A,A'\,)$ to $A \otimes 1 + 1 \otimes A'$.
The inner product on the 2-Hilbert space $\Omega^1(\FR^3, \fru)$ reads as
\begin{equation*}
	\< A, A' \,\>_{\Omega^1(\FR^3, \fru)} = \big\{ g: \FR^3
        \rightarrow \Mat(n{\times}n', \FC) \, \big| \, \dd g + \iu\,
        A'\, g -\iu\,  g\, A = 0 \big\}
\end{equation*}
on objects, whereas on morphisms we find
\begin{equation*}
	\< f, f'\, \>_{\Omega^1(\FR^3, \fru)} = \overline{f} \otimes f'\ .
\end{equation*}
The inner product on the Hilbert space $\< A, A'\, \>_{\Omega^1(\FR^3, \fru)} \in \Hilb$ is given by
\begin{equation*}
	(g,g'\,) \longmapsto \tr \big( g^* \,g'\, \big)\ .
\end{equation*}
This is easily seen to be constant via the calculation
\begin{equation*}
\begin{aligned}
	\dd\, \tr \big( g^* \,g'\, \big) &= \tr \big( \dd (g^*\, g'\, ) \big)\\[4pt]
	&= \tr \big( \dd (g^*\, g'\,) + \iu\, A\, g^* \,g' - \iu\, g^*\,g'\, A \big)\\[4pt]
	&= \tr \big( ( \dd g^* + \iu\, A\, g^* - \iu\, g^* \,A'\, )\,
        g' + g^*\, ( \dd g' + \iu\, A'\, g' - \iu\, g'\, A ) \big)\\[4pt]
	&= 0\,
\end{aligned}
\end{equation*}
by the condition on 2-morphisms.

This category bears a remarkable similarity to the Lie 2-algebras of classical observables in 2-plectic geometry considered by~\cite{Baez:2008bu,Rogers--Thesis} (see~\cite{BSS--HGeoQuan} for further discussion of this point).
It would be interesting to find an extension of this Lie 2-algebra
structure to the entire category of higher functions in the general case, but even an extension to the simpler category $\Omega^1(M,\fru)$ would give valuable insight into the quantised version of the rig category of higher functions.
Understanding the induced noncommutative and nonassociative structure on the complete higher functions would be a huge step towards full higher geometric quantisation.

To finish this section, let us point out a slight modification of the above example which has an interpretation in terms of M-theory.
The M-theory $C$-field serves as an $H$-flux on the 6-dimensional worldvolume of an M5-brane on which open membranes end. On $M=\FR^6=\FR^3{\times}\FR^3$ the $C$-field can be modelled on a trivial bundle gerbe $\CL= \CI_\rho$ on $\FR^3{\times}\FR^3$ with $H$-flux~\cite{Chu:2009iv} $H=2\pi\,(\dd x^1 \wedge \dd x^2 \wedge \dd x^3+\dd x^4 \wedge \dd x^5 \wedge \dd x^6)$, which is a geometric realisation of the 2-plectic form on $\FR^3{\times}\FR^3$ given by the sum of the canonical 2-plectic forms on each of the transverse $\FR^3$ subspaces. The construction of the corresponding prequantum 2-Hilbert space proceeds verbatum to that above, yielding again an equivalence of rig-module categories
\begin{equation*}
	\Gamma(\FR^3 {\times} \FR^3 ,\, \CI_\rho) \cong \Omega^1( \FR^3 {\times} \FR^3,\, \fru)\ .
\end{equation*}

\subsection{M-theory lift of $\FR^2$}
\label{sect:R2xS1}

A bound D2--D4-brane configuration in a $B$-field background of Type~IIA
string theory lifts to a bound M2--M5-brane configuration in a $C$-field
background of M-theory. If the transverse space to the D2-brane worldvolume is $\FR^2$ with its
canonical symplectic form $\omega=\dd x^1\wedge\dd x^2$ fluxed by the
$B$-field, then the transverse space to the membrane worldvolume is
the total space of a circle bundle $M$ over $\FR^2$ with a 2-plectic
form that is fluxed by the $C$-field. Since the base $\FR^2$ is contractible any such fibration is trivial, so that $M = \FR^2 {\times} S^1$ and it comes endowed with the natural 2-plectic form given by $\varpi = \omega\wedge\dd\tau=\dd x^1 \wedge \dd x^2 \wedge \dd \tau$, where $\tau\in[0,1)$ is the coordinate along the M-theory direction $S^1$. Since $M$ is homotopy equivalent to $S^1$, we see that $\rmH^3(M,\RZ) = 0$ and therefore every bundle gerbe on $\FR^2 {\times} S^1$ is trivial.
A geometric representative of the 2-plectic form is given by $\CI_\rho$ with $\rho = \pi \, \epsilon_{ij}\, x^i\, \dd x^j \wedge \dd \tau$.

The similarity to the example of Section~\ref{sect:R3} goes even further.
Isomorphism classes of rank~$n$ hermitean vector bundles (without connection) on $M$ are classified by $[M, \sfB \sfU(n)]$, the homotopy classes of based maps from $M$ to the classifying space $\sfB\sfU(n)$.\footnote{Since $M = \FR^2 {\times} S^1$ is path-connected we can choose an arbitrary base-point. Here, for any topological group $\sfG$, the classifying space $\sfB\sfG\cong|\scB\sfG|$ is the coarse moduli space of the classifying stack $\scB\sfG=(\sfG\rightrightarrows *)$ of principal $\sfG$-bundles.}
We can then use the homotopy equivalence $\FR^2 {\times} S^1 \simeq S^1$ to obtain
\begin{equation*}
 \big[ \FR^2 {\times} S^1,\, \sfB \sfU(n) \big] \cong \big[ S^1,\, \sfB \sfU(n) \big] \cong \big[ \Sigma S^0,\, \sfB \sfU(n) \big] \cong \big[ S^0,\, \Omega \sfB \sfU(n)\big] = \pi_0\big(\Omega \sfB \sfU(n)\big)\ ,
\end{equation*}
where $\Sigma$ denotes the suspension of a topological space.
Using the equivalence $\Omega \sfB \sfG \simeq \sfG$ for any topological group $\sfG$ we find that $[ \FR^2 {\times} S^1,\, \sfB \sfU(n) ]$ is trivial.
Therefore, just as on $\FR^3$, every hermitean vector bundle (without connection) is isomorphic to a trivial hermitean vector bundle.
Hence we have once again an equivalence of rig categories
\begin{equation*}
	\HVBdl^\nabla(M) \cong \Omega^1( \FR^2 {\times} S^1,\, \fru)\ ,
\end{equation*}
and of $\Omega^1( \FR^2 {\times} S^1,\, \fru)$-module categories
\begin{equation*}
	\Gamma(\FR^2 {\times} S^1 ,\, \CI_\rho) \cong \Gamma( \FR^2 {\times} S^1 ,\, \CI_0) \cong \HVBdl^\nabla(M) \cong \Omega^1( \FR^2 {\times} S^1,\, \fru)\ .
\end{equation*}

We would now like to check how this higher prequantisation behaves
under dimensional reduction of M-theory back to string theory along
the $S^1$-direction, wherein the M2--M5-brane system reduces to the
D2--D4-brane system for which ordinary geometric prequantisation
should apply. Dimensional reduction is implemented by integrating over
the $S^1$-fibres. First of all, the 2-plectic form $\varpi = \omega \wedge \dd \tau$ nicely reduces to the symplectic form $\omega = \iota_{\partial_\tau}\, \varpi$ on $\FR^2$.
It is represented by the trivial hermitean line bundle $I\to\FR^2$
with connection $a= \iota_{\partial_\tau}\, \rho = \pi \,
\epsilon_{ij}\, x^i\, \dd x^j$; indeed $\dd a = 2\pi\, \omega$.

Next, given a section $A \in \Omega^1(\FR^2 {\times} S^1,\fru)$, we obtain a
function on $\FR^2$ given by the path-ordered exponential
\begin{equation}\label{eq:Wilsonloop}
(x^1,x^2) \longmapsto \red(A)(x^1,x^2) = \tr \, {\rm P}\,
\exp\Big(\iu\, \int_0^{1}\, A_{|(x^1,x^2,\tau)}(\partial_\tau)\,
\dd \tau \Big) \ .
\end{equation}
The value of $\red(A)$ at $(x^1,x^2) \in \FR^2$ is just the Wilson loop of
the hermitean vector bundle with connection obtained
as the restriction of $(\FR^2{\times} S^1 {\times} \FC^n,\, \dd {+}
\iu\, A)$ to the $S^1$ fibre of $\FR^2 {\times} S^1$ over the point $(x^1,x^2)$.
One can see that 2-isomorphic sections of $\CI_\rho$ reduce to identical functions on $\FR^2$.
Hence the dimensional reduction {descends to 2-isomorphism classes} of sections of $\CI_\rho$.
The fact that the trace converts direct sums into sums and tensor products into products implies that this prescription is compatible with the module action of higher functions in the following sense.
Higher functions on $\FR^2 {\times} S^1$ are given by the category $\Omega^1(\FR^2 {\times} S^1,\fru)$ up to canonical equivalence, since the prequantum bundle gerbe $\CI_\rho$ is topologically trivial.
Therefore the action of higher functions on sections is, in this situation, simply the tensor product of higher functions.
As the holonomy of a tensor product of vector bundles is the tensor
product of the holonomies, we obtain $\red(A \vartriangleleft A'\, ) =
\red(A)\, \red(A'\, )$.

The hermitean 2-bundle metric evaluated on a pair of sections $A,A'
\in \Omega^1(\FR^2 {\times} S^1,\, \fru)$ yields $\frh( A, A'\, ) = -A^\rmt \otimes 1 + 1 \otimes A'$.
Since the occuring 1-forms are self-adjoint we can rewrite this as
$\frh( A, A'\, ) = -\overline{A} \otimes 1 + 1 \otimes A'$, so that
\begin{equation*}
	\red\big(\frh(A, A'\, )\big) = \overline{\red(A)}\, \red(A'\,)
        = h \big( \red(A),\, \red(A'\, ) \big)\ ,
\end{equation*}
where $h$ is the hermitean metric on the trivial hermitean line bundle with connection.
Thus dimensional reduction in this example maps the hermitean 2-bundle
metric on $\CI_\rho$ to the hermitean bundle metric on $(I, \dd {+}
\iu\, a)$.
However, while dimensional reduction works out nicely at the level of
sections and the hermitean structures, it is not evident how to relate
the inner products $\< A,A' \,\>_{\Omega^1(\FR^2 {\times} S^1,\fru)}$
and $\< \, \red(A),\,
\red(A'\, ) \>_{C^\infty(\FR^2,\FC)}$ on the respective categories and spaces of sections.

\subsection{M-theory lifts of lens spaces}
\label{sect:lens}

The next natural example we would like to consider is the case wherein
the transverse space of the D2--D4-brane system is a 2-sphere $S^2$
with its canonical symplectic structure given by the area form;
quantisation in
the T-dual picture then describes the polarisation of open D1-branes
into the usual
fuzzy 2-spheres. In the uplift to M-theory, this
configuration becomes an M2--M5-brane system with transverse space a
3-sphere $S^3$, which is an $S^1$-bundle over $S^2$ via
the Hopf fibration $S^3\to S^2$, and
with its canonical 2-plectic structure given by the volume form.
Quantisation would then describe the polarisation of open M2-branes
into what should be analogously called a `fuzzy 3-sphere', which is
also relevant in non-geometric flux compactifications of closed string
theory originating via T-duality from $S^3$
with geometric $H$-flux (see e.g.~\cite{Blumenhagen:2010hj}); this
quantum geometry has been thus far elusive despite various
attempts. Unfortunately, this example also lies outside our scope of
quantisation, as the cohomology ring of the 3-sphere $S^3$ is
torsion-free so that the interesting prequantum bundle gerbes on $S^3$ which geometrically realise the canonical 2-plectic form do not have non-trivial sections in our framework.\footnote{Generally,
on any manifold $M$, a bundle gerbe $\CL$ admits a section
$(E,\alpha):\CI_0\to\CL$ of finite rank if and only if the Dixmier-Douady class
$\DD(\CL)$ lies in the torsion subgroup ${\rm
  Tor}\big(\rmH^3(M,\RZ)\big)$ of the degree~$3$ cohomology of $M$~\cite{BCMMS--Twisted_K-theory}. In that case, the order of $\DD(\CL)$ is a divisor of the rank of every section $(E,\alpha)$.}

Hence we shall instead consider a modification of this example, which
has non-zero torsion in its third integer cohomology, by taking
quotients of $S^3$; this will provide illuminating modifications of the local 2-Hilbert
spaces $\Omega^1(M,\fru)$ associated to trivial prequantum bundle gerbes.
We employ an explicit construction from~\cite{Johnson:2002tc} of bundle gerbes realising cup product Dixmier-Douady classes, which then allows us to reformulate the category of sections of such bundle gerbes in terms of more familiar equivariant vector bundles.
The twist introduced by the gerbe is thereby translated into a twist between two group actions of $\RZ_p$ and $\RZ$.
This way, we can explore higher geometric structures in terms of ordinary geometry.
To finish the section with, we examine the dimensional reduction of such sections along the circle.
Using the aforementioned view on sections of the bundle gerbe, we see that these nicely reduce to sections of a prequantum line bundle on the lens space which realises its second cohomology.

Geometric set-ups as outlined above arise in string compactifications which are described by an asymmetric orbifold of the $\sfS\sfU(2)$ WZW model by a left-acting cyclic subgroup $\RZ_p\subset\sfU(1)\subset\sfS\sfU(2)$ for $p\in\NN$, see e.g.~\cite{Maldacena:2001ky}. Geometrically, this model describes a closed string propagating on a
3-dimensional lens space $\FL_p=\FL(p;1)=S^3/\RZ_p$, the quotient of the 3-sphere
$S^3\subset\FC^2$ by the free action of the cyclic group acting in the fundamental
representation $\FC^2$ of $\sfS\sfU(2)$. Then $\FL_p$ is connected with
$\pi_1(\FL_p)=\pi_0(\RZ_p)=\RZ_p$ as $S^3$ is simply-connected, and it
is orientable (and spin) if $p\geqslant3$. The generator of the
fundamental group $[f]\in\pi_1(\FL_p)$ may be taken to be any loop
$f:S^1\to \FL_p$ obtained by projecting a path on the universal cover
$q_p:S^3\to \FL_p$ which connects two points on $S^3$ in the same orbit of
the $\RZ_p$-action.

The lens space $\FL_p$ has a CW-complex decomposition with one $k$-cell
$e_k$ in each dimension $k=0,1,2,3$, and a $p$-fold covering map
$e_2\to e_1$~\cite{DavisKirk}. Its singular homology is thus $\rmH_0(\FL_p,\RZ)=\rmH_3(\FL_p,\RZ)=\RZ$, $\rmH_1(\FL_p,\RZ)=\RZ_p$ and $\rmH_2(\FL_p,\RZ)=0$, and hence by the Universal Coefficient Theorem the cohomology ring of $\FL_p$ is given by
\begin{equation*}
\rmH^0(\FL_p,\RZ)=\rmH^3(\FL_p,\RZ)=\RZ \ , \qquad \rmH^2(\FL_p,\RZ)=\RZ_p \ , \qquad \rmH^1(\FL_p,\RZ)=0 \ .
\end{equation*}
The generator of $\rmH^2(\FL_p,\RZ)=\RZ_p$ can be described
geometrically in the following way~\cite{Karoubi}. Let $K\to S^2$ be the hermitean
line bundle canonically associated to the Hopf fibration $S^3\to S^2$
of degree~1. Then $c_1(K)$ is a generator of $\rmH^2(S^2,\RZ)=\RZ$,
and the lens space can be identified as the total space $\FL_p\cong
S(K^{\otimes p})$ of the circle bundle $\Pi_p:S(K^{\otimes p})\to S^2$
of the hermitean line bundle $K^{\otimes p}$ with degree~$p$. With
$J:=\Pi_p^*K$, the Chern class $c_1( J)=\Pi_p^*c_1(K)$ generates $\rmH^2(\FL_p,\RZ)=\RZ_p$.

The third cohomology here is still torsion-free, so we consider an
M-theory lift of the lens space to an oriented $S^1$-bundle $\pi_p:M_p
\to \FL_p$.
The cohomology groups of $M_p$ are related to those of $\FL_p$ through the Gysin exact sequence
\begin{equation*}
\cdots \ \longrightarrow \ \rmH^k(M_p,\RZ) \ \xrightarrow{ \ \pi_{p!} \ } \ \rmH^{k-1}(\FL_p,\RZ) \ \xrightarrow{\smile\,e} \ \rmH^{k+1}(\FL_p,\RZ) \ \xrightarrow{ \ \pi_p^* \ } \ \rmH^{k+1}(M_p,\RZ) \ \longrightarrow \ \cdots
\end{equation*}
where $ \pi_{p!}$ is the Gysin pushforward map on cohomology, $\pi_p^*$ is the usual pullback on cohomology, and $\smile e$ denotes cup product with the Euler class $e\in\rmH^2(\FL_p,\RZ)$ of the fibration. In particular, putting $k=3$ shows that the third cohomology of $M_p$ sits in a short exact sequence
\begin{equation*}
\begin{matrix}
0 \ \longrightarrow & \ \rmH^3(\FL_p,\RZ) \ & \xrightarrow{ \ \pi_p^* \ } \ \rmH^3(M_p,\RZ) \ \xrightarrow{ \ \pi_{p!} \ } & \ \rmH^2(\FL_p,\RZ) \ & \longrightarrow \ 0 \\[-6pt]
 & \parallel & & \parallel & \\[-6pt]
 & \RZ & & \RZ_p & 
\end{matrix}
\end{equation*}
indicating that $\rmH^3(M_p,\RZ)$ can contain torsion. In particular, for
the trivial bundle $M_p= \FL_p{\times} S^1$ with $e=0$ and $\pi_p$ the projection to the first factor, this sequence has a canonical splitting via the K\"unneth theorem to give
\begin{equation*}
\rmH^3(\FL_p{\times} S^1,\RZ)\cong \RZ\oplus \RZ_p \ .
\end{equation*}
Writing $\big[1_{S^1}\big]$ for the generator of $\rmH^1(S^1,\RZ)\cong\rmH^0(S^1,\sfU(1))=\RZ$, the class\footnote{Here and in the following we suppress pullbacks to $\FL_p{\times} S^1$ for ease of notation.}
\begin{equation}\label{eq:deltatorsion}
\delta_p = c_1( J) \, \smile \, \big[1_{S^1}\big]  
\end{equation}
is a generator for the torsion subgroup ${\rm
  Tor}\big(\rmH^3(\FL_p{\times} S^1,\RZ)\big)=\RZ_p$. In the following
we fix a connection $\nabla^{ J}$ on the hermitean line bundle $ J$. Since
the fibration $\Pi_p:S(K^{\otimes p})\to S^2$ has degree~$p$, we may
choose $\nabla^{ J}$ so that its magnetic flux is given by
\begin{equation}\label{eq:FJ}
F^{ J} = 2\pi\, p\, \Pi_p^*(\omega)
\end{equation}
for a normalised symplectic form $\omega$ on $S^2$, i.e. $\int_{S^2}\,
\omega=1$. Then $F^{ J}$ has trivial de~Rham class in $\rmH_{\rm
  dR}^2(\FL_p{\times} S^1)$ because $\rmH_{\rm dR}^2(\FL_p)=0$.

To construct a bundle gerbe $\CL_p$ with connection representing the
torsion class $\delta_p$, we use the
universal covering map $\pi:\FR\to S^1=\FR/\RZ$ to define the $\RZ$-fibration
$1{\times}\pi:Y=\FL_p{\times}\FR\to \FL_p{\times} S^1$. Then the fibre
product $Y^{[2]}$ can
be identified with $(\FL_p{\times} \FR){\times}\RZ$, and the groupoid structure
maps $p_i:Y^{[2]}\rightrightarrows Y$ as $p_0:(x,r,n)\mapsto (x,r)$
and $p_1:(x,r,n)\mapsto (x,r+n)$ for
$(x,r,n)\in\FL_p{\times}\FR{\times}\RZ$. Omitting pullbacks to $\FL_p{\times}\FR$, for the hermitean line bundle
$L\to Y^{[2]}$ we take $ J^{\otimes\RZ}\to
(\FL_p{\times}\FR){\times}\RZ$ with fibres
$ J^{\otimes\RZ}_{|(x,r,n)}= J^{\otimes n}_{|(x,\pi(r))}$; since $ J^{\otimes p}$ is trivialisable as a line bundle without connection, there is a $p$-periodicity
$L_{|(x,r,n+m\,p)}\cong L_{|(x,r,n)}$ for all $m\in\RZ$ reflecting the torsion. The bundle
gerbe multiplication
\begin{equation*}
\mu_{|(x,r,n,m)}:L_{|(x,r,n)}\otimes L_{|(x,r+n,m)}\arisom
L_{|(x,r,n+m)}
\end{equation*}
is given by the canonical isomorphism $ J^{\otimes n}\otimes
 J^{\otimes m}\to  J^{\otimes(n+m)}$. In this way we obtain the topological data $(L,\mu,Y)$ of a bundle
gerbe $\CL_p$ on $\FL_p{\times} S^1$ with Dixmier-Douady invariant
$\DD(\CL_p)=\delta_p$, see e.g.~\cite{Brylinski,Johnson:2002tc}.

It remains to define a connection on the bundle gerbe
$(L,\mu,Y)$. This is achieved by extending this geometric realisation of the cup product
$\smile\, :\rmH^2(\FL_p,\RZ)\otimes_{\RZ}\rmH^1(S^1,\RZ)\to\rmH^3(\FL_p{\times}
S^1,\RZ)$, which defines the cohomology class \eqref{eq:deltatorsion}, to Deligne cohomology. Since the connection $\nabla^{ J}$ on the line bundle
$ J\to\FL_p{\times} S^1$ induces connections on tensor products
$ J^{\otimes n}$, we obtain a connection $\nabla^L$ on the hermitean
line bundle $L$ which is compatible with $\mu$ and has magnetic
flux $F^L\in\Omega^2(\FL_p{\times} \FR{\times}\RZ)$ given by
$F^L_{|(x,r,n)}= n\, F^{ J}_{|(x,\pi(r))}$. The curving
$B\in\Omega^2(\FL_p{\times}\FR)$ given by $B_{|(x,r)}= r\,
F^{ J}_{|(x,\pi(r))}$ easily obeys
$B_{|(x,r+n)}-B_{|(x,r)}=F^L_{|(x,r,n)}$, and by the Bianchi identity
for $\nabla^{ J}$ we obtain $\dd B=\dd r\wedge (1{\times}\pi)^*
F^{ J}=(1{\times}\pi)^*(\dd\tau\wedge F^{ J})$ with $\tau\in[0,1)$, which identifies the $H$-flux
\begin{equation}\label{eq:HfluxLp}
H= F^{ J} \wedge \dd\tau
\end{equation}
having trivial de~Rham class in
$\rmH^3_{\rm dR}(\FL_p{\times} S^1)$ because the magnetic flux
\eqref{eq:FJ} has trivial class $\big[F^{ J}\big]=0$ in $\rmH^2_{\rm
  dR}(\FL_p{\times} S^1)$. In this way we obtain the desired
bundle gerbe with connection $\CL_p=(L,\mu,Y,\nabla^L, B)$. 

The physical meaning of this torsion $H$-flux can be understood as follows. A closed string winding around the non-contractible cycle of the lens space $\FL_p$ can be regarded as an open string on the universal cover $S^3$ (with endpoints in the same orbit of the $\RZ_p$-action), and the pullback $q_p^*F^{ J}$ of the magnetic flux serves as a $B$-field source for open strings on the worldvolume of a fractional D3-brane wrapping $S^3$~\cite{Maldacena:2001ky}. Projecting back thus produces a fundamental string charge (holonomy) in $\FL_p$ which is valued in $\rmH_1(\FL_p,\RZ)=\RZ_p$~\cite{Gukov:1998kn}. In the lift to M-theory, fundamental string winding numbers become membrane wrapping numbers, so that the membrane charge is valued in $\rmH_2(\FL_p{\times}S^1,\RZ)=\RZ_p$~\cite{Acharya:2000gb}. In this way the torsion magnetic flux $F^{ J}$ on $\FL_p$ induces a torsion $H$-flux on $\FL_p{\times}S^1$, regarded as a $C$-field source for open M2-branes on the worldvolume of a fractional M5-brane wrapping the cover $S^3{\times}\FR$ (with the boundary lines of the membranes in the same orbit of the $\RZ_p{\times}\RZ$-action).

This physical picture appears in the description of the prequantum
2-Hilbert space of sections
$\Gamma(\FL_p{\times}S^1,\CL_p)=\BGrb^\nabla(\FL_p{\times}S^1)(\CI_0,\CL_p)$,
which may be described as bundle gerbe modules~\cite{Waldorf--More_morphisms}. A section is a triple $\big(E,\nabla^E,\alpha\big)$, where $E\to\FL_p{\times}\FR$ is a hermitian vector bundle with connection $\nabla^E$ together with parallel isomorphisms
\begin{equation}\label{eq:alphaisom}
\alpha_{|(x,r,n)}:E_{|(x,r+n)} \arisom E_{|(x,r)}\otimes J^{\otimes n}_{|(x,\pi(r))} \ ,
\end{equation}
which by construction are automatically compatible with $\mu$. The
projection $q_p:S^3\to\FL_p$ does not commute with the
$\sfU(1)$-actions, but $q_p^* J=(\Pi_p\circ q_p)^*K\cong
S^3{\times}\FC=I$ which implies that the trivial hermitean line bundle
$S^3{\times}\FC$ can be used as descent data for $ J$: Identifying
$S^{3\,[2]}$ with $S^3{\times}\RZ_p$, the morphism $p_1^*I\to p_0^*I$
is given by $(\hat x,z,\zeta)\mapsto (\hat x\cdot\zeta,
\zeta^{-1}\cdot z)$ for $(\hat x,z,\zeta)\in
S^3{\times}\FC{\times}\RZ_p$. Defining $I_p\to S^3$ to be the trivial
hermitean line bundle with this $\RZ_p$-action, it follows that $ J$ is the descent bundle of
$(I_p,q_p)$. The connection $\nabla^J$ on $J$ is represented in this descent by the connection $\dd+\iu\,\kappa$ on $I_p$ for some $\RZ_p$-invariant 1-form $\kappa$ on $S^3$. Since
$\HVBdl^\nabla$ is a stack, it also follows that the category
$\HVBdl^\nabla(\FL_p{\times}\FR)$ of hermitean vector bundles with
connection on $\FL_p{\times}\FR$ is equivalent to the category
$\HVBdl^\nabla_{\RZ_p}(S^3{\times}\FR)$ of $\RZ_p$-equivariant
hermitean vector bundles with connection on $S^3{\times}\FR$ with
respect to the $\RZ_p$-action $(\hat x,r)\mapsto (\hat x
\cdot\zeta,r)$. Hence $E\to \FL_p{\times}\FR$ lifts to a pair
$\big(\hat E,\phi^{\hat E}\big)$, where $\hat E\to S^3{\times}\FR$ is
a hermitean vector bundle with connection and $\phi^{\hat E}$ is a parallel isomorphism with
\begin{eqnarray*}
\phi^{\hat E}_{\zeta|(\hat x,r)}:\hat E_{|(\hat x\cdot \zeta,r)}\arisom \hat E_{|(\hat x,r)}
\end{eqnarray*}
for all $\zeta\in\RZ_p$. As the $\RZ$-action and the $\RZ_p$-action on
$S^3{\times}\FR$ commute, the isomorphism \eqref{eq:alphaisom} correspondingly lifts to an isomorphism
\begin{equation*}
\hat\alpha_{|(\hat x,r,n)}:\hat E_{|(\hat x,r+n)} \arisom \hat E_{|(\hat x,r)}\otimes I^{\otimes n}_{p|\hat x}
\end{equation*}
of $\RZ_p$-equivariant hermitean vector bundles with connection.

Thus far we have found an equivalence between the category of sections
$\Gamma(\FL_p{\times}S^1,\CL_p)$ and the category whose objects are
triples $\big(\hat E,\phi^{\hat E},\hat\alpha\big)$, with $\big(\hat
E,\phi^{\hat E}\big)\in \HVBdl^\nabla_{\RZ_p}(S^3{\times}\FR)$ and
$\hat\alpha\in\HVBdl^\nabla_{\RZ_p}(S^3{\times}\FR)\big(\hat p_1^*\hat
E,\, \hat p_0^*\hat E\otimes I_p^{\otimes\RZ}\big)$, and morphisms
given by maps $\hat\sigma:(\hat E,\hat\alpha)\to(\hat F,\hat\beta)$ in
$\HVBdl^\nabla_{\RZ_p}(S^3{\times}\FR)\big(\hat E,\, \hat F\big)$ such
that $\hat\beta\circ\hat\sigma=(\hat
\sigma\otimes1)\circ\hat\alpha$. The line bundle $I_p^{\otimes n}$ is
non-trivial as a $\RZ_p$-equivariant bundle without connection: The morphism $m:\hat p_0^*\hat E\otimes I_p^{\otimes\RZ}\to \hat p_0^*\hat E$ defined by
\begin{equation*}
m_{|(\hat x,r)}:\hat E_{|(\hat x,r)}\otimes I^{\otimes n}_{p|\hat
  x}\longrightarrow \hat E_{|(\hat x,r)} \ , \quad \hat e\otimes(\hat
x,z)\longmapsto z\, \hat e \ ,
\end{equation*}
is an isomorphism in $\HVBdl(S^3{\times}\FR)$, but it is not $\RZ_p$-equivariant as
\begin{equation*}
m\big(\phi^{\hat E}_\zeta(\hat e)\otimes(\hat x\cdot\zeta,\, \zeta^{-n}\cdot z)\big)= \zeta^{-n}\cdot z\ \phi^{\hat E}_\zeta(\hat e)= \zeta^{-n}\ \phi^{\hat E}_\zeta\big(m\big(\hat e\otimes(\hat x,z)\big)\big) \ .
\end{equation*}
We may use it to eliminate the additional line bundle with connection $\big(I_p^{\otimes n},\dd+\iu\,n\,\kappa\big)$
in the bundle gerbe action by replacing the section $\big(\hat
E,\nabla^{\hat E},\hat\alpha \big)$ with an ordinary hermitean vector bundle with connection
$\big(\hat E,\nabla^{\hat E}-\iu\,a\big)$, where $a\in\Omega^1(S^3{\times}\FR)$ is defined by $a_{|(\hat x,r)}=r\,\kappa_{|\hat x}$; it is both $\RZ_p$-equivariant, via
the isomorphism $\phi^{\hat E}$ as before, and $\RZ$-equivariant, via
the isomorphism $\psi^{\hat E}=m\circ\hat\alpha$ which is parallel with respect to the modified connection $\nabla^{\hat E}-\iu\,a$. However,
the bundle $\big(\hat E,\phi^{\hat E},\psi^{\hat E}\big)$ is not
$\RZ_p{\times}\RZ$-equivariant, as the fibrewise-linear actions of the groups obey
non-trivial commutation relations
\begin{equation*}
\phi_\zeta^{\hat E}\circ\psi_n^{\hat E} = \zeta^n\ \psi_n^{\hat
  E}\circ \phi_\zeta^{\hat E}
\end{equation*}
for all $\zeta\in\RZ_p$ and $n\in\RZ$. With $\zeta\in S^1$ a primitive
$p$-th root of unity, we call this a \emph{$\zeta$-twisted action} of the
group $\RZ_p{\times}\RZ$, and the triple $\big(\hat E,\phi^{\hat
  E},\psi^{\hat E}\big)$ a \emph{$\RZ_p{\times}_\zeta\,\RZ$-equivariant vector
bundle} (with connection). Morphisms of such bundles still
commute with the group actions, $\hat\sigma\circ\phi^{\hat
  E}=\phi^{\hat E}\circ \hat\sigma$ and $\hat\sigma\circ\psi^{
\hat E}=\psi^{\hat E}\circ \hat\sigma$, and we denote by
$\HVBdl^\nabla_{\RZ{\times}_\zeta\,\RZ_p}(S^3{\times}\FR)$ the category
  whose objects are triples $\big(\hat E,\phi^{\hat E},\psi^{\hat E}\big)$
  and morphisms $\hat\sigma$ as above. Altogether, we have found an
  equivalence of rig categories
\begin{equation*}
\Gamma(\FL_p{\times}S^1,\CL_p)\cong
\HVBdl^\nabla_{\RZ_p{\times}_\zeta\,\RZ}(S^3{\times}\FR) \ .
\end{equation*}

For the category of higher functions, by descent along the surjective
submersion $q_p{\times}\pi:S^3{\times}\FR\to\FL_p{\times}S^1$ we also
obtain an equivalence of rig categories
\begin{equation*}
\HVBdl^\nabla(\FL_p{\times}S^1)\cong\HVBdl^\nabla_{\RZ_p{\times}\RZ}(S^3{\times}\FR)
\ . 
\end{equation*}
Then $\HVBdl^\nabla_{\RZ_p{\times}_\zeta\,\RZ}(S^3{\times}\FR)$ is
naturally a rig module category over
$\HVBdl^\nabla_{\RZ_p{\times}\RZ}(S^3{\times}\FR)$ under the usual tensor
product of hermitean vector bundles with connection on
$S^3{\times}\FR$. In particular, one sees that the obstruction for a
section of $\CL_p$ to descend to a higher function is precisely the
$\zeta$-twisted action of $\RZ_p{\times}\RZ$.

Finally, let us examine the dimensional reduction of the M2-brane
system back to fundamental strings in the asymmetric orbifold of the $\sfS\sfU(2)$ WZW model. The
reduction described in Section~\ref{sect:R2xS1} by integrating over
the $S^1$-fibres can be generalised to any trivial $S^1$-bundle, and 
even to general circle bundles, by using {transgression} techniques
(see~\cite{BSS--HGeoQuan} and references therein). The form of the
class \eqref{eq:deltatorsion} suggests that the bundle gerbe $\CL_p$
on $\FL_p{\times}S^1$
should reduce precisely to the line bundle $J=\Pi_p^*K$ on $\FL_p$. At the level of
integer cohomology, this is implemented by the Gysin pushforward
$\pi_{p!}:\rmH^3(\FL_p{\times}S^1,\RZ)\to \rmH^2(\FL_p,\RZ)$ which
maps the Dixmier-Douady class of $\CL_p$ back to the corresponding
Chern class: $\pi_{p!}(\delta_p)=c_1(J)$. Likewise, the $H$-flux
\eqref{eq:HfluxLp} nicely reduces to the magnetic flux
$F^J=\iota_{\partial_\tau}H$ on $\FL_p$. 

For the reduction of sections, previously we took Wilson loops \eqref{eq:Wilsonloop}
around the $S^1$-fibre of sections of a trivial bundle gerbe. In the
present case our sections are no longer proper vector bundles, but we
can still define Wilson loops along the $S^1$-fibre by regarding them
as Wilson lines along unit length intervals in $\FR$. If $\big(E,\nabla^E,\alpha\big)$ is a
bundle gerbe module for $\CL_p$ of rank $n$ (divisible by $p$), its holonomy around the $S^1$-fibre
then translates to parallel transport in $E$ along $\FR$. Denoting by
$P^E_{(x,r,s)}$ the parallel transport in $E$ along $\FR$ from $(x,r)$
to $(x,r+s)$, we define an isomorphism by the composition
\begin{equation*}
\rmh_{|(x,r)}=\alpha_{|(x,r,1)}\circ P^E_{(x,r,1)}:E_{|(x,r)}\arisom
E_{|(x,r)}\otimes J_{|(x,\pi(r))} \ .
\end{equation*}
We then obtain a section $\red\big(E,\nabla^E,\alpha\big):\FL_p\to J$ by the analogue of the Wilson
loop given by
\begin{equation}\label{eq:redEx}
x\longmapsto \red\big(E,\nabla^E,\alpha\big)(x):= \tr_{J_{|(x,\pi(r))}}\big(\rmh_{|(x,r)}\big) \ \in
\ J_{|x} \ ,
\end{equation}
for any $r\in\FR$, where the vector-valued trace is defined by
\begin{equation*}
\tr_{J_{|(x,\pi(r))}}\big(\rmh_{|(x,r)}\big) := \sum_{i=1}^n\,
\big\langle e_i\otimes k\,,\,
\rmh_{|(x,r)}(e_i)\big\rangle_{E_{|(x,r)}\otimes J_{|x}} \ k
\end{equation*}
for any orthonormal basis $\{e_i\}$ of $E_{|(x,r)}$ and unit vector
$k$ in $J_{|x}$. To see that $\red\big(E,\nabla^E,\alpha\big)(x)$ is well-defined, we use the
fact that $\alpha$ is bicovariantly constant to compute
\begin{equation*}
\begin{aligned}
\tr_{J_{|(x,\pi(r+s))}}\big(\rmh_{|(x,r+s)}\big) &= \sum_{i=1}^n\,
\big\langle e_i\otimes k\,,\,
\rmh_{|(x,r+s)}(e_i)\big\rangle_{E_{|(x,r+s)}\otimes J_{|x}} \ k
\\[4pt]
&= \sum_{i=1}^n\,
\big\langle e_i\otimes k, \big((P^E_{(x,r,s)}\otimes
1_{J_{|(x,\pi(r))}})\circ \rmh_{|(x,r)}\circ
P^E_{(x,r+s,-s)}\big)(e_i)\big\rangle_{E_{|(x,r+s)}\otimes J_{|x}} k \\[4pt]
&= \sum_{i=1}^n\,
\big\langle P^{E\,*}_{(x,r,s)}(e_i)\otimes k\,,\,
\rmh_{|(x,r)} \big( P^E_{(x,r+s,-s)} (e_i) \big)
\big\rangle_{E_{|(x,r)}\otimes J_{|x}}  \ k \ .
\end{aligned}
\end{equation*}
Since the parallel transport operator is unitary, we have
$P^{E\,*}_{(x,r,s)}=P^{E\,-1}_{(x,r,s)}=P^E_{(x,r+s,-s)}$, and since
the connection $\nabla^E$ is hermitean it follows that the set of
vectors $\big\{P^E_{(x,r+s,-s)}(e_i)\big\}$ is an orthonormal basis of
$E_{|(x,r)}$. As the vector-valued trace is independent of the choice of
orthonormal basis, we thus find
\begin{equation*}
\tr_{J_{|(x,\pi(r+s))}}\big(\rmh_{|(x,r+s)}\big) = \tr_{J_{|(x,\pi(r))}}\big(\rmh_{|(x,r)}\big)
\end{equation*}
and hence the definition of the Wilson loop \eqref{eq:redEx} is
independent of $r\in\FR$. Thus we obtain a well-defined map
\begin{equation*}
\red:\Gamma(\FL_p{\times}S^1,\CL_p)\longrightarrow \Gamma(\FL_p,J) \ ,
\end{equation*}
which as previously descends to 2-isomorphism classes of sections of
$\CL_p$, and is compatible with direct sum, tensor product and the
module actions.

\section{Open problems}
\label{sec:open}

In this paper we have shown how a bundle gerbe on a manifold $M$ naturally gives rise to a 2-Hilbert space.
The main ingredients in its construction are the notions of sections of a bundle gerbe, a hermitean 2-bundle metric, and integration of vector bundles with connection over the manifold $M$.
Although this construction is already very well-behaved, it can be improved in several ways.
For instance, it would be desirable to have additive inverses for sections of a bundle gerbe, or more generally for morphisms of bundle gerbes.
A general formalism for ring completions of rig categories has been developed in~\cite{BDRR--Ring_completion}, but it remains to write down explicit models for ring completions of $\HVBdl^\nabla(M)$ and $\Hilb$.
From the perspective of quantisation, additive inverses would be necessary to describe interesting interference phenomena.

It is still unclear whether a more general notion of morphisms of bundle gerbes can be formulated so that even bundle gerbes with non-torsion Dixmier-Douady class admit sections.
For a detailed discussion of this problem, see~\cite{BSS--HGeoQuan}.
Refering to our comments at the end of Section~\ref{sect:2-HSpaces}, it is possible that the topology on the objects of $\Hilb$ has to be weakened in order to allow for such morphisms.

There are also other models for geometric representations of elements of $\rmH^3(M,\RZ)$, as for example Azumaya algebra bundles~\cite{Kapustin:1999di,MMS--Index_of_projective_families,STV--Gerby_Serre-Swan}.\footnote{Alternatively, the group $\rmH^3(M,\RZ)$ classifies principal $\sfP\sfU(\CH)$-bundles on $M$, where $\CH$ is an infinite-dimensional separable Hilbert space, $\sfU(\CH)$ is the group of unitary endomorphisms of $\CH$, and $\sfP\sfU(\CH)=\sfU(\CH)/\sfU(1)$ is an Eilenberg-MacLane space $K(2,\RZ)$, i.e. $\rmH^1(M,\sfP\sfU(\CH))\cong\rmH^3(M,\RZ)$, see e.g.~\cite{Atiyah:2004jv}. By the Serre-Grothendieck theorem, every torsion class in $\rmH^3(M,\RZ)$ can be represented by a principal $\sfP\sfU(n)$-bundle on $M$.}
It might be interesting to relate our constructions to this formalism similarly to~\cite{STV--Gerby_Serre-Swan}.
So far the endomorphism algebra bundles arising from our sections of bundle gerbes as their 2-endomorphisms are all finite-dimensional, but only since we have restricted our considerations to 2-morphisms which are bicovariantly constant.

The question of finding an analogue on $\HVBdl^\nabla(M)$ of the Poisson algebra structure on $C^\infty(M,\FC)$ from geometric quantisation remains open.
This structure might only exist on a suitable ring completion of $\HVBdl^\nabla(M)$, but on the full subcategory of trivial hermitean line bundles with connection a Lie 2-algebra structure has been found in~\cite{Rogers--Thesis} (see~\cite{BSS--HGeoQuan}).
Higher prequantisation would then require finding a representation of that higher Poisson algebra on the category of sections $\Gamma(M,\CL)$ of a prequantum bundle gerbe $\CL$ on $M$.
Finally, a notion of polarisation would be necessary to obtain a physically sensible 2-Hilbert space; such a notion has been proposed in~\cite{Rogers--Thesis}.
It would be desirable to have a good physical example of complete higher quantisation, most probably from string theory or M-theory, against which a mathematical formalism of 2-plectic quantisation could be confronted.

\subsection*{Acknowledgements}

This work was supported in part by the Action MP1405 QSPACE from 
the European Cooperation in Science and Technology (COST). 
The work of S.B. was supported by a James Watt Scholarship. 
The work of R.J.S.\ is supported in part by the Consolidated Grant ST/L000334/1 
from the UK Science and Technology Facilities Council (STFC). 


\bibliographystyle{amsplainurl}

\providecommand{\bysame}{\leavevmode\hbox to3em{\hrulefill}\thinspace}
\providecommand{\MR}{\relax\ifhmode\unskip\space\fi MR }
\providecommand{\MRhref}[2]{%
  \href{http://www.ams.org/mathscinet-getitem?mr=#1}{#2}
}
\providecommand{\href}[2]{#2}

\vspace{0.5cm}

\noindent
(Severin Bunk) \ 
Department of Mathematics and Maxwell Institute for Mathematical
Sciences, Heriot-Watt University, Edinburgh EH14 4AS, UK \ 
{\tt sb11@hw.ac.uk}

\medskip

\noindent
(Richard J.~Szabo) \ 
Department of Mathematics, Maxwell Institute for Mathematical Sciences, and The Higgs Centre for Theoretical Physics,
Heriot-Watt University, Edinburgh EH14 4AS, UK \ 
{\tt r.j.szabo@hw.ac.uk}

\end{document}